\def\etal{{\em et al.~}}

\documentstyle[11pt,aaspp]{article}

\eqsecnum

\slugcomment{To appear in {\em The Astrophysical Journal} 10 July 1996}

\begin{document}

\title{Cosmological Blastwaves and the Intergalactic Medium}

\author{G. Mark Voit\altaffilmark{1}$^,$\altaffilmark{2}}
\affil{Department of Physics and Astronomy, The Johns Hopkins University,
Baltimore, MD 21218}

\altaffiltext{1}{Hubble Fellow}
\altaffiltext{2}{Current Address:  STScI, 3700 San Martin Drive, Baltimore,
MD 21218}

\setcounter{footnote}{0}

\begin{abstract}
Winds from protogalactic starbursts and quasars can drive shocks that
heat, ionize, and enrich the intergalactic medium.  The Sedov-Taylor
solution for point-like explosions adequately describes these blastwaves
early in their development, but as the time since the explosion ($t - t_1$) 
approaches the age of the universe ($t$), 
cosmological effects begin to alter the blastwave's structure and growth rate.
This paper presents an analytical solution for adiabatic 
blastwaves in an expanding universe, valid when the IGM is homogeneous 
and contains only a small fraction of the total mass density ($\Omega_{\rm IGM}
\ll \Omega_0$).  In a flat universe, the solution applies until 
the age of the universe approaches $t_1 \Omega_{\rm IGM}^{-3/2}$, at which
time the self-gravity of the matter associated with the shock compresses the
shocked IGM into a thin shell.  When $\Omega_{\rm IGM} \lesssim 0.03$, 
blastwaves starting after $z \sim 7$ and containing more than 
$10^{57} \, {\rm erg}$ remain adiabatic to relatively low $z$, so this 
solution applies over a wide range of the parameter space relevant to galaxy
formation.  Using this analytical solution, we examine the role protogalactic
explosions might play in determining the state of intergalactic gas at
$z \sim 2 - 4$.  Since much of the initial energy in galaxy-scale blastwaves
is lost through cosmological effects, photoionization by a protogalaxy
is much more efficient than shock ionization.  Shocking the entire IGM by
$z \sim 4$, when it appears to be substantially ionized, is most easily
done with small explosions ($\lesssim 10^{56} \, {\rm erg}$) having a 
high comoving number density ($\gtrsim 1 \, {\rm Mpc^{-3}}$).  Larger-scale
explosions could also fill the entire IGM, but if they did so, they would 
raise the mean metallicity of the IGM well above the levels observed in Ly$\alpha$ clouds.  Since the metal abundances of Ly$\alpha$ clouds are 
small, the metals in these clouds were probably produced in small-scale 
bursts of star formation, rather than in large-scale explosions.  The 
H~I column densities of protogalactic blastwaves are much smaller than 
those of typical Ly$\alpha$ clouds, but interactions between shocks and 
preexisting Ly$\alpha$ clouds can potentially amplify the neutral column
densities of preexisting clouds by a large factor.
\end{abstract}


\section{Introduction}

Gravity initiates the formation of galaxies in the early universe, but 
shortly thereafter, explosions, winds, and ionizing radiation from massive 
stars and active galactic nuclei complicate the hydrodynamics of galaxy
formation and regulate the transformation of gas into stars (see recent
reviews by Bond 1993 and Shapiro 1995).
In the nearby universe we see starbursting galaxies expelling up to
$10^{59} \, {\rm erg}$ of thermal and kinetic energy into 
intergalactic space over the lifetime of the burst (see Heckman, 
Lehnert, \& Armus 1993 for a review).  The explosive output from 
a protogalaxy could be even larger, if most of its stars formed on
the dynamical timescale of the galaxy. 

Once the hot gas from a 
starburst escapes its host it flows into the intergalactic medium 
(IGM), shocking it, and eventually forming a blastwave similar to
a supernova remnant.  Because cosmological effects, such
as the expansion of the universe, eventually govern their behavior,
the explosions that emanate from quasars or protogalaxies have sometimes
been called cosmological blastwaves.  As these shocks pass through 
intergalactic space, they compress the IGM, sweeping it into denser sheets.
Such a picture has naturally motivated suggestions that cosmological
blastwaves are somehow related to the intergalactic Ly$\alpha$ clouds 
whose absorption signatures striate the spectra of high-redshift 
quasars (Ozernoi \& Chernomordik 1978; Ostriker \& Cowie 1981;
Chernomordik \& Ozernoi 1983; Ikeuchi \& Ostriker 1986; Vishniac \& Bust
1987).

Early treatments of cosmological blastwaves posited a baryonic IGM 
that dominated the matter content of the universe (Schwarz, Ostriker,
\& Yahil 1975; Ostriker \& Cowie 1981; Ikeuchi, Tomisaka, \& Ostriker 
1983; Bertschinger 1983).  
Today the most successful cosmological models assume baryons comprise 
a relatively small fraction of the universe's matter content.
Since the older solutions do not apply, cosmological blastwaves
and their impact on the IGM deserve a fresh look from a more 
contemporary perspective.  Computational approaches to such 
``feedback'' during galaxy formation, while rapidly growing 
more sophisticated, are still limited by numerical resolution (e.g.
Cen \& Ostriker 1993; Navarro \& White 1993).
Analytical approaches, like the one taken here, while simplistic 
and idealized, can be more flexible than numerical solutions and 
offer insights complementary to those gained from simulations.  
Here we demonstrate that, in a universe filled primarily 
with collisionless matter, the structure and growth rate of a 
cosmological blastwave can be determined analytically.  This new 
analytical solution turns out to be handy for reassessing the 
significance of protogalactic shocks in the IGM.

The paper is organized as follows. 
Section~2 presents an analytical solution for adiabatic cosmological
blastwaves, valid when the mass density of the IGM is gravitationally 
unimportant.  This solution is directly analogous to the Sedov-Taylor 
solution for adiabatic supernova remnants.  As long as the shock remains 
adiabatic, the postshock density does not exceed four times the ambient intergalactic density until the shell of collisionless matter 
that collects at the shock front becomes self-gravitating.   
Section~3 discusses how ionizing radiation modifies the medium into which
intergalactic shocks propagate, and \S~4 considers the validity of the
fluid approximation in the IGM.  Section~5 explores the global impact of
protogalactic shocks on the temperature and metallicity of the IGM and
derives constraints on the number density and energy scale of blastwave
sources.  Section~6 examines possible relationships between intergalactic
blastwaves and Ly$\alpha$ clouds, and \S~7 recapitulates the important 
results.

\section{Blastwave Development}

Intergalactic shocks can have an enormous impact on the IGM.
Early calculations assuming that the ratio of the IGM mass density to the total mass density of the universe ($\delta_{\rm IGM} \equiv \Omega_{\rm IGM}/
\Omega_0$) was close to unity found that protogalactic blastwaves 
could fill the IGM, heating and ionizing it by 
redshift $z$ of a few (Schwarz \etal 1975).  Subsequent 
treatments attempting to resolve cosmological shock structure in more detail
(Ikeuchi \etal 1983; Bertschinger 1983; Vishniac, Ostriker, \& 
Bertschinger 1985) generally took $\delta_{\rm IGM} = 1$ and found 
that shocks in an IGM dominated universe formed dense shells.  
Ikeuchi \etal (1983) assumed that dense shells would also form when 
$\delta_{\rm IGM} \ll 1$.  At late times this certainly happens.  
An explosion in the IGM at a time $t_1$ since the big bang creates 
a cavity within which the mass density is a fraction $1-\delta_{\rm IGM}$ 
lower than the background density.  In a flat universe ($\Omega = 1$), 
collisionless matter within the cavity gradually overtakes 
the shock front and forms a shell when the age 
of the universe is $\sim t_1 \delta_{\rm IGM}^{-3/2}$ (Bertschinger 1985).  After this time, blastwave evolution asymptotically approaches the usual 
self-similar solutions in which $t_1$ can be neglected, and the 
shocked IGM collapses into a dense layer just behind the shock front 
(Ikeuchi \etal 1983; Bertschinger 1983).  
If $\delta_{\rm IGM} \ll 1$, most of the interesting
interactions of protogalactic shocks with the IGM occur long before the
blastwave reaches the asymptotic self-similar stage.
  
This section presents an analytical solution for adiabatic cosmological 
blastwaves valid when $\delta_{\rm IGM} \ll 1$ and $t \ll t_1 
\delta_{\rm IGM}^{-3/2}$.  Taking advantage of some similarity properties 
of cosmological expansion (see Shandarin 1980 or Shapiro, Struck-Marcell, 
\& Melott 1983), one can transform the cosmological fluid equations into 
a form without an explicit Hubble flow or background density.  When the 
adiabatic index, $\gamma$, of the IGM equals 5/3, the transformed fluid
equations admit a solution analogous to the Sedov-Taylor solution for 
adiabatic supernova remnants in a homogeneous medium.  This solution turns 
out to be widely applicable, since cosmological shocks at $z \lesssim 7$ 
tend to remain adiabatic.  The solution breaks down when either cooling
becomes important or perturbations in the collisionless background
flow become large.

\subsection{Analytical Solution}

Self-similar solutions for cosmological blastwaves can be found 
straightforwardly when the age of the blastwave is much less than 
the current age of the universe or when the age of the blastwave is
infinitesimally close to the current age of the universe (see Ostriker 
\& McKee 1988 for a review).  In these cases, there is only one important
timescale.  The intermediate case has been considered unfit for self-similar treatment because both the current age of the universe ($t$) and the age of 
the blastwave ($t - t_1$) are important.  However, when $\delta_{\rm IGM} 
\ll 1$ a self-similar solution for cosmological blastwaves exists 
if $\gamma = 5/3$ and the blastwave is adiabatic.

To find this solution, we begin with the Newtonian fluid equations, valid as
long as the length scales are much smaller than the horizon of the
universe and the velocities are non-relativistic:
\begin{eqnarray}
     \frac {\partial \rho} {\partial t} + \nabla \rho {\bf u} & = & 0 \\
     \frac {\partial {\bf u}} {\partial t} + ({\bf u} \cdot \nabla) {\bf u}
         & = & - \nabla \Phi - \frac {1} {\rho} \nabla P \\
     \nabla^2 \Phi & = & 4 \pi G (\rho + \rho_G) \\
     \frac {d \epsilon} {dt} & = & \frac {P} {\rho^2} \frac {d \rho} {dt}
          + \rho {\cal H} - \rho^2 \Lambda \\
     \epsilon & = & \frac {1} {\gamma - 1} \frac {P} {\rho} \; \; \; .
\end{eqnarray}
Here, ${\bf r}$ is the coordinate vector, ${\bf u}$ is the corresponding 
velocity, $P$ is the gas pressure, $\epsilon$ is the specific internal
energy, $\rho$ is the IGM mass density, and $\rho_G$ is the mass density
of collisionless (non-baryonic plus condensed baryonic) matter. The 
undisturbed IGM comoves with the Hubble flow.  Deviations from this flow can 
be expressed as a peculiar velocity field ${\bf v} = {\bf u} - H(t) {\bf r}$ under the influence of a perturbed potential field $\phi = \Phi - 2 \pi G \bar{\rho}(t) r^2 / 3$.  In terms of the scale factor $a(t) = [1+z(t)]^{-1}$,
which obeys $\dot{a}/a= H(t)$, the mean mass density is $\bar{\rho}(t) = 
\rho_0 a^{-3}(t)$, where $\rho_0 = \rho_{\rm cr} \Omega_0 = 3 H_0^2 \Omega_0 
/ 8 \pi G$ is the mass density of the universe at the present time ($t_0$).

A transformation originally applied by Shandarin (1980) recasts the 
cosmological fluid equations into a very useful form (see also Shapiro
\etal 1983).  The transformed variables are:
\begin{eqnarray}
     d \hat{t} = a^{-2} dt \; \; \; , \; \; \; \hat{\bf r} = a^{-1} {\bf r}
          \; \; \; , \; \; \; \hat{\bf v} = a {\bf v} \; \; \; , \; \; \;
          \hat{\rho} = a^3 \rho  \nonumber
\end{eqnarray}
\begin{eqnarray}
     \hat{P} = a^5 P \; \; \; , \; \; \; \hat{\epsilon} = a^2 \epsilon
          \; \; \; , \; \; \; \hat{\phi} = a^2 \phi \; \; \; .
\end{eqnarray}
These variables obey the equations
\begin{eqnarray}
     \frac {\partial \hat{\rho}} {\partial \hat{t}} 
          + \hat{\nabla} \hat{\rho} {\bf \hat{v}} & = & 0 \\
     \frac {\partial {\bf \hat{v}}} {\partial \hat{t}} 
          + ({\bf \hat{v}} \cdot \hat{\nabla}) {\bf \hat{v}}
         & = & - \hat{\nabla} \hat{\phi}
            - \frac {1} {\hat{\rho}} \hat{\nabla} \hat{P} \\
     \hat{\nabla}^2 \hat{\phi} & = & 4 \pi G (\hat{\rho} + \hat{\rho}_G
            - \rho_0) a \\
     \frac {d \hat{\epsilon}} {d \hat{t}} 
          & = & \frac {\hat{P}} {\hat{\rho}^2} \frac {d \hat{\rho}} {d\hat{t}}
            + (5 - 3 \gamma) 
             \left( \frac {1} {a} \frac {da} {d \hat{t}} \right) \hat{\epsilon}
            + \hat{\rho} a {\cal H} - \hat{\rho}^2 a^{-2} \Lambda \\
     \hat{\epsilon} 
          & = & \frac {1} {\gamma - 1} \frac {\hat{P}} {\hat{\rho}} \; \; \; .
\end{eqnarray}
The form of these equations closely follows that of the initial set except 
that $\hat{r}$ and $\hat{v}$ now represent comoving positions and peculiar velocities.

When $\delta_{\rm IGM} \ll 1$, perturbations in the IGM have little effect
on the underlying gravitational potential.  If we also assume that the 
underlying collisionless matter is distributed effectively homogeneously 
on the scale of the blastwave, terms containing $\hat{\phi}$ can be ignored.
In an adiabatic shock, the heating and cooling terms are negligible. The only 
term remaining that depends on cosmic time is the second term in
the energy equation, which vanishes if $\gamma = 5/3$.  This term accounts
for the change in $\hat{\epsilon}$ that arises when $\epsilon$ does not vary
as $a^{-2}$.  In a monatomic gas, all the internal energy is in translational
degrees of freedom.  These atomic motions can be viewed as peculiar velocities
that redshift away like $a^{-1}$, so that $\epsilon \propto a^{-2}$ in the
absence of heating, cooling, and noncosmological work.  Energy contained in 
rotational and other internal degrees of freedom does not redshift away and 
adds to $\hat{\epsilon}$ as the universe ages.  Thus, in the transformed 
frame, the second term in the energy equation heats the gas when 
$\gamma > 5/3$.

The above equations are formally identical to the adiabatic fluid equations
without cosmological and gravitational terms when ${\cal H} = \Lambda = \phi
= 0$ and $\gamma = 5/3$.  Since the time transformation involves only the time
derivative, we are free to define $\hat{t} = 0$ to coincide with the cosmic
time $t_1$ at which a galactic explosion (quasar or starburst) introduces an 
energy $E_0 = a^{-2} \hat{E}_0$ into the IGM.  Solving for $\hat{t}(t)$ 
requires knowing $a(t)$ and thus $\Omega_0$:
\begin{equation}
     \hat{t} = \frac {2} {\Omega_0 H_0} 
                 [(1 + \Omega_0 z_1)^{1/2} - (1 + \Omega_0 z)^{1/2}]
                \; \; . 
\end{equation} 
The standard Sedov self-similar solution (Sedov 1959, 1993) then satisfies 
the fluid equations, giving a shock radius of
\begin{eqnarray}
     \hat{R}_s = \left[ \frac {\xi \hat{E}_0} {\hat{\rho}_{\rm IGM}} 
                    \right]^{1/5} \hat{t}^{2/5} \; \; ,
\end{eqnarray}
where $\xi = 2.026$, $\hat{\rho}_{\rm IGM} = \rho_{\rm cr} \Omega_{\rm IGM}$ 
and $\Omega_{\rm IGM} \equiv \Omega_0 \delta_{\rm IGM}$.  At early times, 
$\hat{t} \approx (1+z_1)^2 (t - t_1)$, and the shock radius in physical
coordinates is
\begin{eqnarray}
     R_s = \left[ \frac {\xi E_0} {\rho_0 \Omega_{\rm IGM} (1+z_1)^3}
           \right]^{1/5} (t - t_1)^{2/5} \; \; ,
\end{eqnarray}
as expected.  At large times, $\hat{t}$ asymptotically approaches a constant,
so $\hat{R}_s$ approaches a constant comoving radius.  
The complete solution for the shock radius is
\begin{eqnarray}
     R_s & = & \left[ \frac {32 \pi \xi G E_0} {3 H_0^4 \Omega_{\rm IGM}} 
                \frac {(1+\Omega_0 z_1)} {\Omega_0^2 (1+z_1)^2} \right]^{1/5}
               \left[ 1 - \left( \frac {1+\Omega_0 z} {1+\Omega_0 z_1}
                \right)^{1/2} \right]^{2/5}
                (1+z)^{-1} \\
      ~  & = & (7.5 \, {\rm Mpc}) \, 
                \left[ \frac {1+\Omega_0 z_1} {\Omega_0^2 (1+z_1)^2}
                   \right]^{1/5}
                E_{61}^{1/5} h_{50}^{-4/5} \Omega_{\rm IGM}^{-1/5}
               \left[ 1 - \left( \frac {1+\Omega_0 z} {1+\Omega_0 z_1}
                \right)^{1/2} \right]^{2/5}
                (1+z)^{-1} \; \; , \nonumber
\end{eqnarray}
where $E_{61} = E_0 / (10^{61} \, {\rm erg})$.
This solution is also valid when $\Omega_{\rm IGM} = \Omega_0 \ll z_1^{-1}$
because the gravitation of the collisionless matter is insignificant 
in this limit.

Since $\Omega_{\rm IGM} \ll 1$, the asymptotic radii of cosmological 
shocks can exceed 10~Mpc in comoving coordinates.  Figure~1 shows how 
quickly the shock radius and shocked mass approach their asymptotic 
values in a flat universe.  Voit (1994), solving for cosmological 
blastwave evolution in the thin-shell approximation, obtained an 
identical rate of convergence to the asymptotic solution.  At the 
asymptotic radius, the blastwave has spent all the input energy in 
lifting the shocked IGM out of the gravitational potential of the 
collisionless matter interior to the shock.  
In the thin-shell approximation, the blastwave must lift all of the 
shocked IGM to the maximum radius, so the asymptotic radius in the 
thin-shell case is slightly smaller than in the exact solution.

\subsection{Cooling}

The onset of efficient cooling ends the adiabatic stage of blastwave 
development.  Inverse Compton scattering of hot electrons off the cosmic
microwave background cools cosmological shocks effectively at $z > 8
h_{50}^{2/5} - 1$ in a flat universe.  Radiative cooling is more important 
at lower redshifts but does not cool the shocked IGM very quickly.  
The cooling function $\Lambda(T)$ at temperature $T$ 
in a low-metallicity plasma is of order $10^{-23} \, {\rm erg \, cm^3 \, 
s^{-1}}$ for $10^5 \, {\rm K} < T < 10^8 \, {\rm K}$ (e.g. Sutherland \& 
Dopita 1993; Schmutzler \& Tscharnuter 1993).  Just behind a strong shock 
front, the hydrogen number density $n_{\rm H}$ is four times the ambient 
IGM hydrogen density and the gas temperature is $T = 3 \mu v_s^2 / 16 k$, 
where $v_s$ is the shock velocity relative to the Hubble flow and $\mu$
is the mean mass per particle.  When the postshock cooling time $t_c = 
5kT/n_{\rm H}\Lambda(T)$ is smaller than $t$, the current age of the 
universe, a cosmological shock is not adiabatic.  If $\Omega_0 = 1$, 
cosmological shocks cool radiatively for
\begin{eqnarray}
     v_s & \lesssim & \left[ \frac {32} {15 \pi} 
                      \frac {H_0 \Omega_{\rm IGM} \Lambda} {G m_p^2}
               \right]^{1/2} (1+z)^{3/4} \\
     ~   & \lesssim & (77 \, {\rm km \, s^{-1}}) 
               \Omega_{\rm IGM}^{1/2}
               \left( \frac {\Lambda} {10^{-23} \, {\rm erg \, cm^3 \, s^{-1}}}
                      \right)^{1/2}
               h_{50}^{1/2} \, (1+z)^{3/4} \, . \nonumber
\end{eqnarray}
If $\Omega_0 \ll z^{-1}$, the criterion for radiative cooling becomes
\begin{eqnarray}
     v_s & \lesssim & \left[ \frac {16} {5 \pi} 
                      \frac {H_0 \Omega_{\rm IGM} \Lambda} {G m_p^2}
               \right]^{1/2} (1+z) \\
     ~   & \lesssim & (96 \, {\rm km \, s^{-1}}) 
               \Omega_{\rm IGM}^{1/2}
               \left( \frac {\Lambda} {10^{-23} \, {\rm erg \, cm^3 \, s^{-1}}}
                      \right)^{1/2}
               h_{50}^{1/2} \, (1+z) \, . \nonumber
\end{eqnarray}
Ionization losses can become important before radiative losses do.
Blastwaves propagating into a neutral medium expend much of their energy
collisionally ionizing the neutral gas when $v_s \sim 40 - 
80 \, {\rm km \, s^{-1}}$.  However, ultraviolet photons from the
protogalaxy or quasar driving the blastwave are likely to preionize the
IGM into which the shock propagates (\S~3).

The analytical solution of \S~2.1 straightforwardly gives the 
velocity at the leading edge of an adiabatic shock:
\begin{eqnarray}
     v_s & = & \frac {2} {5} \left[ 
                \frac {\pi \xi G H_0 E_0} {3 \Omega_{\rm IGM}} 
               \frac {\Omega_0^3 (1+z_1)^3} {(1+\Omega_0 z_1)^{3/2}} 
                   \right]^{1/5}
               \left[ 1 - \left( \frac {1+\Omega_0 z} {1+\Omega_0 z_1}
                  \right)^{1/2} \right]^{-3/5}
               \left( \frac {1+z} {1+z_1} \right) \\
      ~  & = & (75 \, {\rm km \, s^{-1}}) 
               \left[ \frac {\Omega_0^2 (1+z_1)^2} {1+\Omega_0 z_1}
                 \right]^{3/10} 
                E_{61}^{1/5} h_{50}^{1/5} \Omega_{\rm IGM}^{-1/5}
               \left[ 1- \left( \frac {1+\Omega_0 z} {1+\Omega_0 z_1}
                \right)^{1/2}\right]^{-3/5} 
               \left( \frac {1+z} {1+z_1} \right) \; \; . \nonumber
\end{eqnarray}
Note that $v_s(z)$ depends only on $\Omega_0$, $z_1$, and the parameter
combination $E_{61} h_{50} \Omega_{\rm IGM}^{-1}$.
Figure~2 shows how $v_s$ varies with $z$ for various values of $E_{61}
h_{50} \Omega_{\rm IGM}^{-1}$ and several different initial redshifts,
assuming $\Omega_0 = 1$.  The vertical line indicates where inverse Compton
scattering is marginally effective in cooling intergalactic shocks.  
The effectiveness of radiative cooling depends on the density and
metallicity of the ambient IGM.  Metallicities of Ly$\alpha$ clouds are
probably $\lesssim 10^{-2}$ times solar (Tytler \& Fan 1994; Cowie \etal
1995).  At these levels of enrichment, $\Lambda(T)$ 
is nearly equal to the cooling function for a metal-free plasma.
Solid lines across the lower portion of Figure~2 show where $t_c \approx t$, 
according to the cooling functions of Sutherland \& Dopita (1993),
for a metallicity of $10^{-2}$ times solar and $\Omega_{\rm IGM} = 0.01$ 
and $0.03$.  The shocks described here do not cool 
via radiative processes as long as $v_s(z)$ lies to the right of 
the Compton cooling region and above the radiatively cooling
region.  For a wide range of parameters, cosmological blastwaves 
are adiabatic.  The analytical solution derived here is therefore 
valid in many interesting cases.

Once cooling becomes efficient, the blastwave grows more slowly than in 
the adiabatic case.  If the shock cools radiatively, but the interior does
not, then $\hat{E}_0 \propto \hat{R}_s^{-2}$ and $\hat{R}_s \propto
\hat{t}^{2/7}$.  This case is directly analogous to the pressure-driven
snowplow solution for supernova remnants.  If both the shock and the interior
of the blastwave radiate efficiently, then blastwave momentum is constant
in the transformed frame ($\hat{E}_0 \propto \hat{R}_s \hat{t}^{-1}$) and
$\hat{R}_s \propto \hat{t}^{1/4}$, as in the momentum-conserving snowplow
solution for supernova remnants.  For more on supernova-remnant solutions
see Ostriker \& McKee (1988).

Cooling reduces the ultimate radius of a cosmological blastwave and the
total amount of mass it can shock.  Let $\hat{R}_{\rm asy}$ be the asymptotic 
comoving radius of an adiabatic blastwave, and define $\eta \equiv 
\hat{R}_s / \hat{R}_{\rm asy}$.  Suppose that at $\eta = \eta_c$ and 
$\hat{t} = \hat{t}_c$ the blastwave begins to cool efficiently and
obeys $\hat{R}_s = \hat{R}_{\rm asy} \eta_c (\hat{t}/\hat{t}_c)^\beta$
thereafter.  The new asymptotic comoving radius of this
blastwave, after cooling, is $\hat{R}_{\rm asy} \eta_c^{1-5\beta/2}$.
Relative to the adiabatic case, the asymptotic shocked mass is 
$\eta_c^{6/7}$ times smaller when $\beta = 2/7$ and $\eta_c^{9/8}$
times smaller when $\beta = 1/4$.

\subsection{Gravitation of the Shocked Shell}

Shock compression of the IGM creates a radial perturbation in the
gravitational potential that can modify the interior structure of a 
cosmological blastwave.
In transformed variables, the surface density of the shocked IGM shell 
is $\hat{\sigma} = H_0^2 \Omega_{\rm IGM} \hat{R}_s / 8 \pi G$.
Note that the form of equation~(2-9) implies $\hat{G} = G(1+z)^{-1}$.
Behind the shock, the gravitational pressure scale height in the
transformed system is 
\begin{equation}
     \hat{\lambda}_g \sim \frac {1+z} {H_0^2 \Omega_{\rm IGM}}
              \frac {\hat{v}_s^2} {\hat{R}_s} \, .
\end{equation}
Thus, the gravitational pressure scale height in physical 
space is
\begin{equation}
     \lambda_g  \sim  \frac {1} {25 \, \Omega_{\rm IGM}}
          \left( \frac {1+z} {1+z_1} \right) \, R_s \; ,
\end{equation}
when $\Omega = 1$.
In the absence of gravity, the density scale height inside an adiabatic
blastwave is $\sim R_s / 12$, so self-gravity does not alter the radial
density structure of the blastwave significantly until $\lambda_g \sim
R_s/12$ at $t \sim t_1 
\Omega_{\rm IGM}^{-3/2}$.  Gravitational forces modify the radial 
pressure profile of an adiabatic cosmological
blastwave somewhat earlier.  Without gravity, the gas pressure at $r <
0.8 \, R_s$ plateaus at $\sim 0.4 \, P(R_s)$ (Sedov 1959, 1993).  
The gravitational potential of the shocked shell steepens this plateau
when $\lambda_g \sim R_s$.   
We can always neglect the self-gravity of an adiabatic shock 
when $\Omega_{\rm IGM} \ll \Omega \ll 1$.

\subsection{Fragmentation}

Density perturbations in the swept-up shell of gas behind a cosmological
shock can become gravitationally unstable, leading to fragmentation of
the shell and possibly the formation of Ly$\alpha$ clouds or galaxies
(e.g. Ostriker \& Cowie 1981).  
However, when $\delta_{\rm IGM} \ll 1$, gas shells behind adiabatic 
shocks do not fragment until the response of the collisionless 
matter starts to govern the dynamics.  Formation of condensed 
gaseous structures through shock fragmentation alone thus requires either 
shocks that cool radiatively or $\delta_{\rm IGM} \approx 1$.

Let us examine the development of density perturbations in the 
transformed variables when $\Omega_{\rm IGM} \ll \Omega_0 = 1$ 
and the shock is adiabatic.
A density perturbation of wavelength $\hat{\lambda}$ in a gas shell 
with a characteristic sound speed $\hat{c}_s$ grows if the sound
crossing time $\hat{\lambda} \hat{c}_s^{-1}$ exceeds the free-fall
time $(\hat{\lambda} / \hat{G} \hat{\sigma})^{1/2}$
(Ostriker \& Cowie 1981; Vishniac 1983; Voit 1988).
Thus, the minimum timescale, in transformed coordinates, for postshock 
IGM density perturbations to grow gravitationally is $\hat{t}_g \sim 
\hat{c}_s / \hat{G} \hat{\sigma}$.  While the shock remains adiabatic,
$\hat{c}_s \approx 0.16 \, \hat{R}_s / \hat{t}$, and 
$\hat{t}_G \sim 9 (1+z) \Omega_{\rm IGM}^{-1} t_0^2 \hat{t}^{-1}$. 
Postshock perturbations do not grow significantly until
$\hat{t} \sim 3 (1+z)^{1/2} \Omega_{\rm IGM}^{-1/2} t_0$, which
corresponds to the epoch in physical time when $t \sim t_1 
\Omega_{\rm IGM}^{-3/2}$.  At this late point in the development of 
the blastwave, we can no longer ignore the peculiar flow induced in
the collisionless matter as the IGM is swept up.

This stability analysis can be formulated more rigorously following 
Vishniac (1983), who examined the behavior of spherical blastwaves 
in the thin-shell approximation.  
Applying Vishniac's formalism to the gravitational stability problem
in transformed coordinates yields a similar growth time, 
$\hat{t}_g \approx \hat{c}_s / \pi \hat{G} \hat{\sigma}$, and
the same qualitative result.  If $\delta_{\rm IGM} \ll 1$,
adiabatic cosmological shocks do not fragment gravitationally
until the response of collisionless matter to the blastwave
begins to drive the shock front.

Radiative cooling, when it finally becomes a significant energy sink,
reduces $\hat{c}_s$, allowing gravitational instabilities to develop 
earlier, but isothermal shocks are subject to a disruptive 
dynamical overstability that operates much more 
quickly than gravitational collapse (Vishniac 1983;
Bertschinger 1986).  Ripples in the shocked shell induce transverse flows
of gas from the leading regions of the shell to the trailing regions.
As the surface density of the trailing regions grows, the trailing regions
begin to overtake the leading regions, and the process reverses.  When 
the overstability becomes nonlinear, weak shocks damp the transverse flows, staving off complete disruption of the main shock front (Grun \etal 1991; 
Mac~Low \& Norman 1994).  Gravitational instabilities therefore proceed
to grow roughly as expected.

\subsection{Comparisons with Other Solutions}

Since the pioneering work of Schwarz \etal (1975), several investigators 
have presented numerical and analytical solutions for cosmological 
blastwaves.  Many of these treatments assume $\delta_{\rm IGM} = 1$
and are not valid when $\Omega_{\rm IGM} \ll \Omega_0$.
At late times ($t \gg t_1 \Omega_{\rm IGM}^{-3/2}$) in a flat universe
filled with collisionless matter,
\begin{eqnarray}
     R_s & = & 1.89 \, (E_0 G)^{1/5} t^{4/5} \\
         & = & (4.4 \, {\rm Mpc}) E_{61}^{1/5} (1+z)^{-6/5} \, .
           \nonumber
\end{eqnarray}
(Bertschinger 1985).
If $\Omega_{\rm IGM} \lesssim 0.1$, this solution does not yet apply 
to blastwaves beginning after the era of Compton cooling ($z_1 < 7$),
because the collisionless component has not yet responded fully to the 
motion of the IGM.  Several workers have formulated approximate expressions 
for $R_s(t)$ in the important regime where $t \ll t_1 \Omega_{\rm IGM}^{-3/2}$ 
and $\Omega_{\rm IGM} \ll \Omega_0 = 1$.  Here we compare these expressions 
with the exact solution derived in \S~2.1.

\subsubsection{Shock Radius}

Adiabatic cosmological blastwaves start as spherical Sedov shocks and
then asymptotically approach a constant comoving radius before the 
collisionless matter drives $R_s$ to the late-time solution. 
Ozernoy \& Chernomordik (1978) constructed an approximate expression for
$R_s$ by matching a Sedov solution at early times to the solution of
Schwarz et al. (1975) at late times.  This approximation is poor because
it presumes that the collisionless component responds to the blastwave
on a timescale $\sim t_1$.  Vishniac \& Bust (1987) introduced a correction 
factor to the Sedov solution that reproduces, with reasonable accuracy, 
the behavior of $R_s(z)$ in numerical models of adiabatic cosmological 
shocks (Vishniac \etal 1985)
when $\Omega_{\rm IGM} \ll \Omega = 1$.  Figure~3 shows how their
approximation (VB87) compares with the exact solution of \S~2.1 (V95)
and the late-time self-similar solution of Bertschinger (B85)
for $\Omega_{\rm IGM} = 0.1$ and 0.01.  The corrected Sedov solution remains
within 10\% of the exact solution until $t \sim 100 t_1$, regardless of 
$\Omega_{\rm IGM}$.  After this time, the slope of VB87 is similar
to B85.  The point of departure of VB87 from V95 should be later for 
$\Omega_{\rm IGM} = 10^{-2}$, indicating that the VB87 correction 
factor was derived from models with $\Omega_{\rm IGM} \sim 0.1$.
Ostriker \& McKee (1988) formulated an expression for blastwave 
expansion that can be compared more directly with the solution
of \S~2.1.  They constructed an approximate solution for adiabatic blastwaves 
when $\Omega_{\rm IGM} \ll \Omega_0 = 1$ by matching a Sedov-Taylor
solution at early times to a comoving shell solution at late times.    
The approximation of Ostriker \& McKee (1988), labeled OM88 in Figure~3, 
agrees well with the exact solution until $(1+z_1)/(1+z) \sim 4$
and ultimately reaches a somewhat smaller comoving radius.

The asymptotic comoving radius ($\hat{R}_{\rm asy}$) of a blastwave 
solution in a flat universe dominated by collisionless matter depends 
upon $E_0$, $\Omega_{\rm IGM}$, $z_1$, $H_0$, and also on the form of 
$R_s(t)$.  When the other parameters are equal, solutions that expand 
more slowly reach larger values of $\hat{R}_{\rm asy}$.  This is why 
the V95 and OM88 solutions eventually differ.  
To see how the shock expansion rate affects $\hat{R}_{\rm asy}$, 
consider how the initial energy $E_0$ shifts from the IGM to the
collisionless component, assuming that the shock sweeps the IGM into a thin
shell.  As an IGM shell of mass $M$ overtakes a collisionless shell of mass $dM_d$, the gravitational energy of the collisionless shell increases by 
$GM \, dM_d / R_s$ and the gravitational energy of the IGM shell decreases by
the same amount.  In terms of $\eta(x) \equiv \hat{R}_s / \hat{R}_{\rm asy}$ 
and $x \equiv (1+z)/(1+z_1)$, we can write the energy transfer rate as
\begin{equation}
     \frac {dE} {dx} = \frac {3 \, H_0^4 \hat{R}_{\rm asy}^5 
                               \Omega_{\rm IGM} (1+z_1)}  {4 \, G} 
                         \, x \eta^4 \eta^{\prime} \; .
\end{equation}
The total amount of energy transferred approaches $E_0$ as $x \rightarrow 0$;
hence, 
\begin{equation}
     \hat{R}_{\rm asy} = \left[ \frac {4 \, G E_0} 
                                {3 \, H_0^4 \Omega_{\rm IGM} (1+z_1)}
                                \right]^{1/5}
                   \left( - \int_0^1 x \eta^4 \eta^{\prime} \, dx
                          \right)^{-1/5} \; .
\end{equation}
From \S~2.1 we have $\eta(x) = (1 - x^{1/2})^{2/5}$ when $\Omega_0 = 1$, 
which leads to
\begin{equation}
     \hat{R}_{\rm asy} = \left[ \frac {40 \, G E_0} 
                                {H_0^4 \Omega_{\rm IGM} (1+z_1)}
                                \right]^{1/5} \; .
\end{equation}
This is identical to the result obtained by Voit (1994) in the thin-shell
approximation.  The value of $\hat{R}_{\rm asy}$ from \S~2.1 is 11\% larger 
than this because the IGM density in the exact solution falls off more 
gradually behind the shock front, slightly reducing the amount of 
gravitational energy associated with a given asymptotic radius.  
Note that blastwaves starting at later times expand to larger comoving 
radii.  Ostriker \& McKee (1988) took
\begin{equation}
   \eta(x) = \left[ ( 1 - 0.66 \, x^{3/2}) (1 - x^{3/2})^2 \right]^{1/5}
         \; ,
\end{equation}
which yields an asymptotic radius in the thin-shell approximation 16\% 
smaller than that in equation (27).  The actual asymptotic radius of 
the OM88 solution is 20\% smaller than the value of $\hat{R}_{\rm asy}$ 
from \S~2.1.  Figure~3 shows that the OM88 form for $\eta(x)$ increases 
more rapidly at early times, transferring the initial energy 
to the dark matter sooner, when the energy needed to achieve a 
given comoving radius is larger.

\subsubsection{Shock Velocity}

Although the blastwave radii of the approximate solutions do not differ
dramatically, the predicted shock velocities can be quite different at
late times.  Figure~4 displays the shock velocities corresponding to the
approximations illustrated in Figure~3.  Because the OM88 approximation
approaches its asymptotic radius so quickly, the shock velocity drops
off rapidly beyond $t \sim 4 t_1$.  In this approximation the shock
cools prematurely.  The VB87 approximation is more 
accurate and remains close to V95 until $t \sim 30 t_1$ ($\sim 
\Omega_{\rm IGM}^{-3/2} t_1$ for $\Omega_{\rm IGM} = 0.1$.)

\subsubsection{Postshock Density}

Immediately behind the shock front the gas density is four times the 
current IGM density.  Because the solution of \S~2.1 is directly analogous
to the Sedov solution, the postshock density drops as $r$ decreases.
Gravitational forces can compress the shell further, but the compression
does not become significant until $t \sim \Omega_{\rm IGM}^{-3/2} t_1$
(\S~2.3).  Ikeuchi \etal (1981) argue that explosions in the IGM should
eventually produce dense shells, but postshock densities in their numerical
calculations do not exceed $4 \rho_{\rm IGM}$ until $t \gtrsim 10 t_1$.

\section{Ionization}

At $z < 5$, the intergalactic medium appears to be highly ionized 
(Gunn \& Peterson 1965; Steidel \& Sargent 1987; Webb \etal 1992; 
Giallongo \etal 1994).   Blastwaves at these redshifts can expand 
adiabatically until radiative cooling limits their growth (\S~2.2).  
When the IGM is not yet fully ionized, shocks initially propagate 
into a cosmological H~II region created by ionizing radiation 
from the source of the blastwave, presumably a quasar or a protogalactic 
starburst.  If the blastwave energy exceeds the ionizing output of the 
source, the shock can potentially enlarge the ionized cavity.  
Shocks reaching the neutral IGM at speeds greater than $\sim 50 \, 
{\rm km \, s^{-1}}$ collisionally ionize additional gas.
This section determines the conditions under which shock heating adds to 
the total mass of ionized gas.

Ultraviolet radiation from a quasar or starburst in the early universe 
quickly ionizes the surrounding IGM.  Analytical models for the expansion of 
cosmological H~II regions have been developed by Donahue \& Shull (1987),
Shapiro \& Giroux (1987), and Madau \& Meiksin (1991).  The recombination
time in a uniform IGM is $(6.3 \times 10^{10} \, {\rm yr}) (1+z)^{-3} 
\Omega_{\rm IGM}^{-1} h_{50}^{-2} T_4^{0.8}$, where $T_4 = T / 10^4 \, 
{\rm K}$.  Thus, at $z < 7$, ionized intergalactic gas with 
$\Omega_{\rm IGM} < 0.1$ remains predominantly ionized, and 
a burst of ionizing radiation produces a permanent H~II region.  
Although the neutral fraction in this region does not necessarily 
remain small enough to satisfy existing constraints on the total 
optical depth of intergalactic Ly$\alpha$, it does permit blastwaves
to propagate at speeds slower than $80 \, {\rm km \, s^{-1}}$ without
suffering heavy ionization losses.  If the central source emits 
$(10^{61} \, {\rm erg})E_{I61}$ in ionizing photons of average energy
$(13.6 \, {\rm eV}) E_{\rm Ryd}$, the total gas mass of the
ionized region is
\begin{equation}
     M_{I} = (5.4 \times 10^{14}\, M_\odot) \, E_{I61} 
                       E_{\rm Ryd}^{-1} \; ,
\end{equation}
assuming primordial abundances.  This mass corresponds to a comoving radius
$\sim (12.3 \, {\rm Mpc}) E_{I61}^{1/3} E_{\rm Ryd}^{-1/3} 
h_{50}^{-2/3} \Omega_{\rm IGM}^{-1/3}$.  The H~II region grows much more
rapidly than any hydrodynamic disturbance until the ionizing source shuts
off and grows slowly thereafter as diffuse recombination radiation 
continues to provide ionizing photons.  After the ionizing episode, 
the warm ionized gas can drive a $10-20 \, {\rm km \, s^{-1}}$ shock into 
the neighboring neutral IGM.  Madau \& Meiksin (1991) employed a special 
case of the transformation described in \S~2.1 in their analytical models of
these shocks. 

Collisional ionization adds to the ionized mass when the blastwave 
energy exceeds the radiative ionizing energy and the energy of the 
blastwave is relatively small.  The mass of shocked gas inside an 
adiabatic cosmological blastwave is $M_{\rm asy} \eta^3$, where
\begin{eqnarray}
     M_{\rm asy} & = & \frac {H_0^2 \Omega_{\rm IGM}} {2 \, G}
                       \left[ 
                        \frac {32 \pi \xi G E_0} 
                              {3 H_0^4 \Omega_{\rm IGM}}
                        \frac {(1+\Omega_0 z_1)} {\Omega_0^2 (1+z_1)^2}
                       \right]^{3/5}  \\
                 & = & (1.2 \times 10^{14} \, M_\odot) 
                        \left[ \frac {1 + \Omega_0 z_1} {\Omega_0^2 (1+z_1)^2} 
                         \right]^{3/5}                         
                         E_{61}^{3/5} h_{50}^{-2/5} \Omega_{\rm IGM}^{2/5}
                         \;  \nonumber
\end{eqnarray}
in a flat universe.  Thus, an adiabatic blastwave eventually expands 
beyond the preionized region if
\begin{equation}
     \frac {E_{I61}} {E_{61}} \: < \: 0.22 \, E_{\rm Ryd} 
           \left[ E_{61} h_{50} \Omega_{\rm IGM}^{-1}
            \frac {\Omega_0^3 (1+z_1)^3} {(1+\Omega_0 z_1)^{3/2}} 
           \right]^{-2/5} \; .
\end{equation}
Blastwaves generated by large-scale energy sources ($E_{61} \gtrsim 1$)
remain within the preionized region unless the initial burst of ionizing
radiation is quite weak.  Smaller-scale blastwaves can escape the preionized
region more easily.  The thermal energy density immediately behind a 
$\sim 50 \, {\rm km \, s^{-1}}$ shock is just enough to ionize all the 
incoming hydrogen.  Below this threshold, the postshock gas remains 
mostly neutral.  Let $\eta_I$ be the dimensionless comoving radius at 
which $v_s = 50 \, {\rm km \, s^{-1}}$, so that $M_{\rm asy} \eta_I^3$
is the total mass of gas a cosmological blastwave can ionize.
Figures~5 and 6 show how $\eta_I$ and $M_{\rm asy} \eta_I^3$ depend on 
the parameter combination $E_{61} h_{50} (1+z_1)^{3/2} 
\Omega_{\rm IGM}^{-1}$ in a flat universe.  Blastwaves in an open
universe behave similarly.

When $\eta_I \ll 1$, a large proportion of the blastwave's energy can go 
into ionization of hydrogen.  Small-scale blastwaves propagating into 
a neutral IGM lose significant energy to ionization when $v_s \sim 
50 - 80 \, {\rm km \, s^{-1}}$ and cool via Ly$\alpha$ when $v_s 
\sim 10 - 50 \, {\rm km \, s^{-1}}$, as long as $\Omega_{\rm IGM} 
\gtrsim 0.01$.  The blastwave switches from an adiabatic solution 
to a pressure-driven snowplow solution at $\eta = \eta_I$, so  
the total mass within the blastwave asymptotically approaches 
$M_{\rm asy} \eta_I^{6/7}$ (\S~2.2).  In blastwaves with $\eta_I$ 
approaching unity, most of the initial energy goes into the collisionless 
component when $\Omega_0 \approx 1$ or into kinetic energy that redshifts
away when $\Omega_0 \ll 1$, leaving little available for ionization.  
Consequently, large-scale explosions are quite inefficient at 
ionizing the IGM.

Around quasars, photoionization exceeds collisional ionization.
Although some quasars might expel winds with luminosities comparable 
to their radiative luminosities (Voit, Weymann, \& Korista 1993), 
constraints on the distortion of the microwave background 
spectrum imply that the total energy in quasar 
blastwaves cannot exceed the total radiative output of quasars by 
a large factor (Voit 1994).  Since the integrated radiative output of a 
typical quasar is $\sim 10^{61} \, {\rm erg}$, quasar blastwaves 
remain within their preionized cavities (see Fig. 6).

The ionizing properties of protogalaxies are harder to constrain because 
it is difficult to measure the fraction of ionizing photons that escape.
In a starburst, the ratio of hydrodynamic to photoionizing output,
integrated over the burst, can be as large as 0.3 (Leitherer \& Heckman
1995).  If the galaxy itself absorbs $> 70$\% of the ionizing photons while
allowing all of the hydrodynamic energy to escape, collisional ionization 
could, in principle, exceed photoionization in the surrounding IGM.
For this to happen, the total magnitude of the protogalactic starburst would 
have to be quite small.  If $\Omega_{\rm IGM} = 0.1$, $h_{50}=1$, 
and $E_{I61}/E_{\rm Ryd} E_{61} = 0.3$, a blastwave beginning 
at $z_1 = 5$ would not augment the photoionized cavity unless 
$E_{61} < 10^{-4.7}$, implying $< 10^5$ supernovae.

\section{Comments on the Fluid Approximation}

Most investigations of intergalactic hydrodynamics, including this one,
assume that the fluid equations adequately describe the behavior of the 
IGM, yet the mean free paths of fast particles in the IGM are not
necessarily small.  Coulomb collisions with electrons in a $10^4$~K 
plasma stop a proton moving at $10^8 \, v_8 \, {\rm km \, s^{-1}}$ over 
an electron column density $\sim 10^{16} \, v_8^4 \, {\rm cm^{-2}}$ 
(Spitzer 1962).  This column corresponds to a comoving distance $\sim 
(0.1 \, {\rm Mpc}) (\Omega_{\rm IGM} / 10^{-2})^{-1} v_8^4$.  In a hot 
plasma at $10^7 \, T_7 \, {\rm K}$, the column a proton traverses before 
Coulomb collisions significantly deflect it is $\sim 10^{17} 
\, T_7^2 \, {\rm cm}^{-2}$ (Spitzer 1962), corresponding to a 
comoving distance $\sim (1 \, {\rm Mpc})(\Omega_{\rm IGM} / 
10^{-2})^{-1} T_7^2$.  These distances are similar to the length scales 
that emerge from cosmological blastwave solutions. 

An intergalactic magnetic field of $10^{-20} \, B_{20} \, {\rm G}$ 
confines protons moving at $10^8 \, v_8 \, {\rm K}$ to within 
a gyroradius of $0.3 \, v_8 B_{20}^{-1} \, {\rm Mpc}$, so a 
primordial field exceeding $10^{-19} \, {\rm G}$ would 
restore the fluid approximation on intergalactic scales.  The actual
strength of the intergalactic magnetic field is hard to constrain (Kronberg 1994).  Because the current Galactic field ($\sim 10^{-6} \, {\rm G}$)
must arise from an initially small seed field, some investigators have
evaluated the primordial field strengths required by various models for
Galactic field generation.  Microwave background scattering 
inhibits the motions of electrons in rotating protogalaxies, causing the 
electrons to lag behind the ions.  This weak current generates a small
($10^{-21} \, {\rm G}$) intergalactic field (Mishustin \& Ruzmaikin 1972).
Galactic dynamo action can potentially amplify a seed field this small
by a factor $\sim e^{N_{\rm rot}} \sim 10^{15}$, where $N_{\rm rot} \sim 
30$ is the number of times the Galaxy has rotated (Parker 1979).  
Kulsrud \& Anderson (1992) have criticized this picture, arguing
that is it very difficult for a galactic dynamo to create the current
field from a very small seed field (but see Field 1995).  
The Galactic field would then have to arise from a much larger 
($\sim 10^{-13} \, {\rm G}$) mean primordial field, amplifed primarily 
by the initial compression of the protogalactic gas.\footnote{Certain 
scenarios for inflation in the early universe can produce 
primordial fields as high as $10^{-9} \, {\rm G}$ (Ratra 1992).}
Uncertain detections of Faraday rotation in radio-selected QSOs behind 
damped Ly$\alpha$ systems at $z \approx 2$ suggest that early 
galactic disks had fields $\sim 10^{-6} \, {\rm G}$ when $N_{\rm rot} 
\ll 30$ and would support the view that the intergalactic field must be 
$\gg 10^{-21} \, {\rm G}$ (Wolfe, Lanzetta, \& Oren 1992).  

In interstellar and interplanetary shocks, magnetic fields and plasma 
turbulence transfer energy between particles more efficiently than Coulomb 
collisions do, allowing postshock fluid structures to develop 
on length scales much shorter than the Coulomb mean free path (Draine 
\& McKee 1993).  Even in the absence of a magnetic field, electrostatic
shocks mediated by plasma turbulence can thermalize the incoming particle
energy in a distance similar to the Mach number times the Debye length
(Tidman \& Krall 1971).  Since such shocks are quite thin, even in the
IGM, the radii of cosmological blastwaves are still well defined and
we can safely use the fluid equations and strong-shock jump conditions
to describe how they propagate.

\section{Global Effects}

At redshift $z \approx 4$ the intergalactic medium is already highly ionized,
and Ly$\alpha$-absorbing clouds have formed.
If cosmological blastwaves are responsible for maintaining the high
ionization of the IGM or for producing Ly$\alpha$ clouds, they must have
filled intergalactic space by this time.  Hot electrons inside such 
blastwaves scatter microwave background photons, distorting the
microwave background spectrum (Weymann 1965, 1966; Sunyaev \& 
Zeldovich 1980).  Limits on this distortion restrict the total blastwave 
energy density in the early universe.  
Metallicity limits further constrain the energy density in supernova 
blastwaves, if the metal abundances of Ly$\alpha$ clouds reflect
the average metallicity of the IGM.  The metallicity constraint
allows adiabatic blastwaves beginning at $z_1 < 7$
to shock most of the IGM by $z \approx 4$ only if the energy per blastwave
is small and the comoving number density of the blastwave sources 
is high ($\sim 1 \, {\rm Mpc}^{-3}$).  Thus, the metals in Ly$\alpha$
clouds were probably produced in small-scale starbursts ($< 10^{56}
\, {\rm erg}$) rather than in large-scale protogalactic superwinds ($> 10^{57} \, {\rm erg}$).  This section outlines these constraints in more detail.

\subsection{Comptonization}

The COBE satellite has shown that the cosmic microwave background spectrum
deviates little from a pure blackbody.  This lack of deviation can be used
to limit the amount of energy injected into the IGM by cosmological blastwaves.
Stars with a standard initial mass function release a fraction 
$\epsilon_{\rm SN} \sim 5 \times 10^{-6}$ of their initial rest mass energy
in the form of kinetic energy from supernovae.  The mean thermal energy
density introduced into the IGM by supernovae in a star forming episode at
$z_1$ is then
\begin{equation}
     U_1(z_1) = (1.5 \times 10^{-14} \, {\rm erg \, cm^{-3}}) (1+z_1)^3
          h_{50}^2 f_{\rm esc} \Omega_*  \; \; ,
\end{equation}
where $f_{\rm esc}$ is the fraction of the supernova energy that escapes
a star-forming galaxy and $\Omega_*$ is the mass density of stars formed
at $z_1$, in units of the critical density.\footnote{About 72\% of the 
initial energy passes into thermal energy as the blastwave develops
(Sedov 1959).}  Since $\hat{E}_0$ is constant in adiabatic cosmological
blastwaves, $U_1(z) = U_1(z_1) [(1+z)/(1+z_1)]^5$.  Shocks in a 
low-density, preionized IGM remain adiabatic until $z$ has changed
significantly (Figure~2).  The amount of $y$-distortion expected from
such a star-forming episode is 
\begin{equation}
     y_1 \approx (2.2 \times 10^{-5}) \, h_{50} f_{\rm esc}
          \Omega_* [(1+z_1)^{3/2} - (1+z_1)^{-2}]  \; \; ,
\end{equation}
assuming $\Omega = 1$. 
COBE finds $y < 2.5 \times 10^{-5}$ at the 2$\sigma$ level (Mather \etal 
1994), implying $\Omega_*  h_{50} f_{\rm esc} < 0.05 - 0.08$ for
$z_1 = 7 - 5$.  At $z_1 > 8 h_{50}^{2/5} - 1$ in a flat universe, 
Compton scattering transfers all the supernova energy to the 
microwave background, resulting in the limit $\Omega_* f_{\rm esc} 
< 2.8 \times 10^{-3} \, (1+z_1)$.

Fluctuations in the numbers of blastwaves along different lines of sight will cause the brightness of the microwave background to vary on small angular scales (Hogan 1984; Yoshioka \& Ikeuchi 1987, 1988), but these fluctuations are relatively insignificant as long as the blastwave energies do not greatly 
exceed $10^{61} \, {\rm erg}$.  A single blastwave distorts the microwave
background by an amount 
\begin{equation}
     \Delta y \sim (9 \times 10^{-10})
           \left[ \frac {\Omega_0^2 (1+z_1)^2} {1 + \Omega_0 z_1} \right]^{2/5}
            \left( \frac {1+z} {1+z_1} \right)^2 
             E_{61}^{3/5} h_{50}^{8/5} \Omega_{\rm IGM}^{2/5} \eta^{-2}
            \; \; \; ,
\end{equation}
which reduces the background's brightness temperature by $\Delta T / T 
\approx -2 \Delta y$.  The temperature fluctuations induced by individual 
blastwaves are unobservable unless $\eta$ is very small or $E_{61}$ is 
very large. 

Brightness temperature fluctuations from an ensemble of blastwaves depend 
upon how many blasts lie along a typical line of sight.  The cross section 
of a blastwave is $\propto \eta^2$, but the amount of time a blastwave 
spends at a size $\eta \ll 1$ is $\propto \eta^{5/2}$.  The number 
of blastwaves of size $\eta$ along a line of sight is therefore ${\cal N}_\eta 
\propto \eta^{9/2}$.  Large blastwaves contribute most heavily to $y$, so
${\cal N}_\eta(\eta \sim 1) \sim y/\Delta y$ and
\begin{equation}
     {\cal N}_\eta < (3 \times 10^4)
          \left[ \frac {\Omega_0^2 (1+z_1)^2} {1 + \Omega_0 z_1} \right]^{-2/5}
            \left( \frac {1+z_1} {1+z} \right)^2 
             E_{61}^{-3/5} h_{50}^{-8/5} \Omega_{\rm IGM}^{-2/5} \eta^{9/2}
            \; \; \; .
\end{equation}
When there are many blasts along a line of sight, the $y$ fluctuations from blasts of size $\eta$ are $\sim {\cal N}_\eta^{1/2} \Delta y \propto 
\eta^{1/4}$.  Blasts with $\eta \sim 1$ thus dominate the temperature
fluctuations, which are of order
\begin{equation}
     \frac {\Delta T} {T} < (3 \times 10^{-7})
            \left[ \frac {\Omega_0^2 (1+z_1)^2} {1 + \Omega_0 z_1} \right]^{1/5}
            \left( \frac {1+z} {1+z_1} \right) 
             E_{61}^{3/10} h_{50}^{4/5} \Omega_{\rm IGM}^{1/5}
            \; \; \; .
\end{equation}
The temperature fluctuations induced by galaxy-scale explosions, which cannot
exceed $\sim 10^{-7} \, E_{61}^{3/10}$ for $z_1 \sim 5$ and $\Omega_{\rm IGM} 
\sim 0.01$, are presently unobservable.

Brightness temperature fluctuations imprinted by individual young 
($\eta \ll 1$) blastwaves can exceed $\Delta T / T \sim 10^{-6}$, but these
are rare.  When $\Omega_{\rm IGM} \sim 0.01$, explosions with $\eta < 0.02
E_{61}^{3/10}$ can produce temperature fluctuations at this level.  The
angular sizes of such blasts at $z \sim 4$ in a flat universe would be 
$\sim 5 \, E_{61}^{1/2} \, {\rm arcsec}$, and they would cover a fraction 
$< (3 \times 10^{-3}) E_{61}^{3/4}$ of the sky.

\subsection{Metallicity}

Supernova-driven winds from galaxies carry metals into intergalactic space, 
and the fraction of the metal-bearing supernova ejecta that escapes a 
starbursting galaxy should be $\sim f_{\rm esc}$.  Since stars formed 
with a normal initial mass function eventually expel about 1\% of the 
input mass in the form of metals, a star-formation episode enriches the 
IGM to a mean level $\sim f_{\rm esc} \Omega_* / \Omega_{\rm IGM}$ 
times solar metallicity.  If the enrichments of Ly$\alpha$ clouds 
reflect average IGM abundances, then upper limits on the intergalactic metallicity constrain the energy input from supernova-driven blastwaves.
Lu (1991) derived Ly$\alpha$-cloud metallicities $\sim 10^{-3}$ times 
solar from stacked C~IV spectra of clouds with neutral columns
exceeding $N_{\rm H \, I} > 10^{15} \, {\rm cm^{-2}}$.  Recent Keck 
spectra of smaller clouds (Cowie \etal 1995) 
now indicate metallicities $\sim 10^{-2} - 10^{-2.5}$ times solar.
Similar metallicities in the global IGM would imply $f_{\rm esc} 
\Omega_* / \Omega_{\rm IGM} \lesssim 10^{-2}$.

This constraint assumes that metals escaping from protogalaxies are 
evenly mixed into the IGM and Ly$\alpha$ clouds.  The assumption of
even mixing is most justifiable if protogalactic blastwaves filled the 
IGM before Ly$\alpha$ clouds condensed.  If metal-bearing blasts 
encompassed only a small fraction of the IGM by the time Ly$\alpha$
clouds formed, the variance in intergalactic metallicities would be high.  
Enriched clouds would have high metallicities, while the others would be 
metal-free.  Tytler \& Fan (1994) determined that Ly$\alpha$ clouds with
metallicities $\ll 10^{-2}$ times solar must be extremely rare, and 
Cowie \etal (1995) find a surprising consistency in cloud metallicities.
While it is possible that the clouds are self-enriched by associated star
formation (e.g. Cowie \etal 1995; Madau \& Shull 1995) and do not reflect 
the metallicity of the IGM, they would still provide an upper bound on 
IGM enrichment at the time of cloud formation.  The metallicity constraint 
would not apply if Ly$\alpha$ clouds formed {\em before} protogalactic 
blastwaves reheated the IGM and avoided mixing with the hotter
metal-rich gas as it swept past.  Nevertheless, further explorations 
of metallicity variations in Ly$\alpha$ clouds might provide very useful
constraints on the star-formation history of the universe. 

\subsection{Porosity}

Blastwaves initiated after $z = 7$ can fill the IGM by $z = 4$, while
still satisfying the above constraints on microwave background distortion
and metallicity, as long as the typical explosion size is relatively
small.  Blastwaves of characteristic energy scale $E_0$ stemming from 
star formation at $z_1$ have a comoving number density 
$\hat{n}_b = (6.3 \times 10^{-2} \, {\rm Mpc^{-3}}) \Omega_* f_{\rm esc}
E_{61}$.  Multiplying this number density by the 
typical blastwave volume gives a porosity parameter
\begin{equation}
     Q = 111 \, \Omega_* f_{\rm esc} E_{61}^{-2/5} h_{50}^{-2/5}
          \Omega_{\rm IGM}^{-3/5} (1+z_1)^{-3/5} \eta^3 
\end{equation}
analogous to the porosity parameter for supernova remnants in the Galactic
interstellar medium (Cox \& Smith 1974).  Blastwave sources distributed 
randomly in space shock a fraction $\sim 1 - e^{-Q}$ of the total
IGM volume.  Shocks from clustered sources merge early, increasing the
effective $E_0$ and lowering the porosity, so $Q$ can be considered an
upper limit on the true porosity.  
Writing $Q$ in terms of the typical blastwave shock velocity $v_s$ yields a
particularly illuminating expression:
\begin{equation}
     Q = \left( \frac {f_{\rm esc} \Omega_*} {\Omega_{\rm IGM}} \right)
          \left( \frac {v_s} {790 \, {\rm km \, s^{-1}}} \right)^{-2}
            x^2 \; \; , 
\end{equation}
where $x = (1+z)/(1+z_1)$.  The initial energy input decays adiabatically 
($\propto x^2$), and $Q$ approaches unity when this energy
has been distributed evenly throughout the IGM.  The Comptonization
limit on $f_{\rm esc} \Omega_*$ at $z_1 = 7$ permits $Q > 1$ when 
$v_s < (180 \, {\rm km \, s^{-1}}) \Omega_{\rm IGM}^{-1/2} h_{50}^{-3/2} x$.  The more stringent metallicity limit permits $Q > 1$ when $v_s < (79 
\, {\rm km \, s^{-1}}) x$.  

More generally, let $Z_{-2}$ be the IGM metallicity in units of 
$10^{-2}$ times the solar metallicity.  Spreading the blastwave energy 
throughout the IGM gives an average IGM temperature $\sim (2 \times 10^5 
\, {\rm K}) Z_{-2} x^2$, if there are no ionization or radiative losses.  
Requiring $Q > 1$ at $z = 4$ for $z_1 = 7$ in a flat universe then 
implies $E_0 < (4 \times 10^{57} \, {\rm erg}) 
Q^{-5/2} Z_{-2}^{5/2} h_{50}^{-1} \Omega_{\rm IGM}$ and
$\hat{n}_{b} > (1.6 \,  {\rm Mpc}^{-3}) h_{50}^3 Z_{-2}^{-3/2}$.  
This density is much larger than the number density of $L_*$ galaxies 
and is comparable to the number density of Ly$\alpha$ clouds (\S~6.1).
Figures~7 and 8 illustrate these constraints graphically for 
$\Omega_{\rm IGM} = 0.01$ and 0.03.
The enrichment levels of Ly$\alpha$ clouds thus suggest that the metals 
they contain originated in relatively small-scale starbursts close 
to the clouds themselves or in individual population~3 supernovae rather 
than in enormous protogalactic starbursts.
 
Quasar blastwaves do not necessarily enrich the IGM with metals, thus
they are not bound by the metallicity limit.  The Comptonization limit,
however, still applies.  Because the number density of quasars 
is low, the energy scale of their blastwaves must be high, if these 
blastwaves are to fill the universe quickly.  The energy 
density needed for quasar blastwaves to give $Q > 1$ by $z \sim 4$ 
violates the COBE limits on $y$-distortion (Voit 1994; see also 
Figures~7 and 8).  Apparently, quasars are too rare to shock-heat the
entire IGM before Ly$\alpha$ clouds appear.

\section{Blastwaves and Ly$\alpha$ Clouds}

Although we have been assuming the IGM is homogeneous, intergalactic 
space is obviously inhomogeneous.  The IGM is filled with clouds
responsible for the numerous Ly$\alpha$ absorption lines, known 
collectively as the Ly$\alpha$ forest, that litter the spectra of 
high-redshift quasars.  The Ly$\alpha$ clouds remain puzzling.  Many 
authors have speculated that these clouds might be linked somehow 
to cosmological blastwaves.  Some have suggested that the 
absorption lines come from the intergalactic shock fronts themselves
(e.g. Chernomordik \& Ozernoy 1983).  They could also be condensed
fragments of primordial blastwaves, now pressure-confined by the IGM
(e.g. Ostriker \& Cowie 1981; Ikeuchi \& Ostriker 1986; Vishniac \& Bust 
1987), gas that has settled into shallow mini-halos of dark matter 
(Rees 1986; Bond, Szalay, \& Silk 1988), gas collecting at caustics 
in the dark-matter velocity field (McGill 1990), or pancakes of shocked
gas where that velocity field converges (Cen \etal 1994).  This section
explores whether cosmological blastwaves produce observable Ly$\alpha$
absorption lines and examines how these shocks interact with preexisting 
intergalactic clouds.  We find that the postshock H~I column densities 
($N_{\rm H \, I}$) 
in the ambient IGM are much smaller than those of intergalactic Ly$\alpha$
clouds, but blastwaves that interact with preexisting clouds can amplify
the Ly$\alpha$ equivalent widths of these clouds significantly.

\subsection{Ly$\alpha$ Cloud Characteristics}

The joint distribution function describing the number of Ly$\alpha$
clouds ${\cal N}$ per redshift interval per column density interval 
can be written
\begin{equation}
     \frac {\partial^2 {\cal N}} {\partial N_{\rm H \, I} \, \partial z} 
          = {\cal N}_0 N_{14}^{-\beta} (1+z)^\gamma \; \; ,
\end{equation}
where $N_{14} = N_{\rm H \, I} / 10^{14} \, {\rm cm^{-2}}$.  Marginally 
opaque clouds ($N_{14} \sim 1$) absorb most of the photons when $1 < 
\beta < 2$. For this reason, clouds with $N_{\rm H \, I} \sim 10^{14} 
\, {\rm cm^{-2}}$ are considered typical, even though a power law 
distribution of column densities fits the data.
Press \& Rybicki (1993) find ${\cal N}_0 \approx 2.6$ and $\beta = 1.43 
\pm 0.04$ in a column density range at least as large as $10^{13} \, 
{\rm cm^{-2}} \lesssim N_{\rm H \, I} \lesssim 10^{16} \, {\rm cm^{-2}}$ for
$\gamma = 2.46$ at redshifts $2.5 \lesssim z \lesssim 4$ (Press, Rybicki, 
\& Schneider 1993).\footnote{The redshift evolution 
parameter $\gamma$ is still not well constrained; current values 
range from 1.89 (Bechtold 1994) to 2.75 (Lu, Wolfe, \& Turnshek 1991).}  
The average comoving distance between clouds with $N_{\rm H \, I} > 
10^{14} \, {\rm cm^{-2}}$ is then $\sim (7 \, {\rm Mpc}) [(1+2.5 \Omega_0)/3.5]^{-1/2} h_{50}^{-1}$ at $z = 2.5$ and $\sim 
(1.7 \, {\rm Mpc}) [(1+4 \Omega_0)/5]^{-1/2} h_{50}^{-1}$ at $z = 4$.  

Near the redshift of the background quasar, the number density of 
Ly$\alpha$ absorption lines decreases (Weymann, Carswell, \& Smith 1981;
Murdoch \etal 1986).  This ``proximity effect'' presumably occurs 
where the ionizing radiation field of the quasar overwhelms the 
ambient ionizing field.  The mean intensity of ionizing radiation 
at $z \approx 3$ inferred from the proximity effect is $(10^{-21}
\, {\rm erg \, cm^{-2} \, s^{-1} \, Hz^{-1} \, ster^{-1}}) J_{21}$,
where $J_{21} \sim 1 - 3$ (Bajtlik, Duncan, \& Ostriker 1988; 
Bechtold 1994).
In photoionization equilibrium, the fractional abundance of atomic
hydrogen in an intergalactic cloud is
\begin{equation}
     f_{\rm H^0} \approx 2.8 \times 10^{-7} \, (1+z)^3 \, D_{\rm cl} 
          \Omega_{\rm IGM} h_{50}^2 J_{21}^{-1} T_4^{-0.8} \; \; ,
\end{equation}
where $D_{\rm cl}$ is the ratio of the cloud density to the IGM density and
$T_4$ is the cloud temperature in units of $10^4$~K, giving a total column
density for a typical cloud of
\begin{equation}
     N_{\rm H} = (3.6 \times 10^{20} \, {\rm cm^{-2}}) \, (1+z)^{-3} \,
          N_{14} D_{\rm cl}^{-1} \Omega_{\rm IGM}^{-1} 
          h_{50}^{-2} T_4^{0.8} J_{21}
           \; \; .
\end{equation}
Recent observations of the quasar pair Q1343+2640A, B, indicate that the 
transverse cloud radius $r_{\rm cl} = (200 \, {\rm kpc}) h_{50}^{-1} 
r_{200}$ for clouds with $N_{14} \sim 1$ lies in the range $80 \, 
h_{50}^{-1} \, {\rm kpc} < r_{\rm cl} < 560 \, h_{50}^{-1} \, 
{\rm kpc}$ at $z \approx 1.8$ (Bechtold \etal 1994; see also 
Dinshaw \etal 1994).  This result implies
\begin{eqnarray} 
     D_{\rm cl} & \sim & 0.8 \, N_{14}^{1/2} \Omega_{\rm IGM}^{-1} 
                 h_{50}^{-3/2} T_4^{0.4} J_{21}^{1/2} r_{200}^{-1/2}   \\
     N_{\rm H} & \sim & (2 \times 10^{19} \, {\rm cm^{-2}}) N_{14}^{1/2}
          h_{50}^{-1/2} T_4^{0.4} J_{21}^{1/2} r_{200}^{1/2}
\end{eqnarray}
at $z \sim 2$, if the clouds are roughly spherical.  Sheetlike clouds
would have a larger $D_{\rm cl}$ and smaller $N_{\rm H}$.  
Assuming spherical clouds, Bechtold \etal (1994) derive a comoving 
number density $\sim (2 \, {\rm Mpc}^{-3}) h_{50}^3 r_{200}^{-2}$ 
for clouds of $N_{14} \sim 1$ at $z \sim 2$.
Perhaps coincidentally, this density is similar to the comoving 
number density of blastwave sources needed to enrich the entire IGM 
to $10^{-2}$ times solar by this redshift (\S~5.3).

\subsection{IGM Ionization Constraints}

We see a Ly$\alpha$ forest because the spaces between Ly$\alpha$ clouds
are either gas free or highly ionized.  Gunn \& Peterson (1965) inferred
$f_0 \Omega_{\rm IGM} \equiv \Omega_{\rm H~I} < 6 \times 10^{-7} \, 
h_{50}^{-1}$ from the lack of continuous absorption at $z \approx 2$. 
Observations at the wavelength of He~II Ly$\alpha$ (304~\AA) toward a quasar 
at $z = 3.286$ show that intergalactic He~II absorbing gas fills velocity 
space at $z \sim 4$ (Jakobsen \etal 1994).  Although this observation 
appears to confirm that the IGM has a spatially continuous component, the
cumulative He~II absorption of many very thin velocity-broadened
Ly$\alpha$ clouds ($N_{\rm H \, I} \gtrsim 10^{-12} \, {\rm cm^{-2}}$) could also produce the observed signal (Madau \& Meiksin 1994).
Current limits on $\Omega_{\rm H \, I} = 10^{-8} \, \Omega_{8}$ in
a spatially continuous component are $\Omega_8 < 4.3 \, h_{50}^{-1}$
at $z \approx 2.5$ (Steidel \& Sargent 1987) and $\Omega_8 < 2.5 \, 
h_{50}^{-1}$ at $z \approx 4$ (Webb \etal 1992; Giallongo \etal 1994).
Photoionization equilibrium thus implies $\Omega_{\rm IGM} < 0.016 \, 
h_{50}^{-1} T_4^{0.4} J_{21}^{1/2} \Omega_8^{1/2}$ at $z = 4.1$.

Alternatively, the IGM could be collisionally ionized by early supernovae
(Tegmark, Silk, \& Evrard 1993),
but if the metallicity of the IGM is similar to that of Ly$\alpha$ 
clouds, photoionization is likely to be more important.  The mean 
temperature in an IGM heated by supernova-driven blastwaves is 
$\lesssim (2 \times 10^5 \, {\rm K}) Z_{-2}$ (\S~5.3), and  
pure collisional ionization at $2 \times 10^5 \, {\rm K}$ requires
$\Omega_{\rm IGM} < 3.6 \times 10^{-3} \, \Omega_8$ for Ly$\alpha$ to
be transparent at the observed levels, a condition more stringent 
than the photoionization condition.  
An elevated temperature does, however, reduce the IGM recombination rate, 
easing the photoionization constraint on $\Omega_{\rm IGM}$ (e.g. Shapiro
1995).

\pagebreak

\subsection{Shocked Shells as Ly$\alpha$ Clouds}

In principle, a sufficiently large compression of the IGM in a shock along
the line of sight to a quasar could produce an observable Ly$\alpha$ feature
(Ozernoy \& Chernomordik 1978; Chernomordik \& Ozernoy 1983).  This can 
happen if $\Omega _{\rm IGM} \sim 1$, but since $\Omega_{\rm IGM} 
\lesssim 10^{-2}$, as the ionization constraints indicate, the H~I 
column densities of cosmological blastwaves in the ambient IGM are 
much smaller than typical Ly$\alpha$-cloud column densities 
($N_{\rm H \, I} \sim 10^{14} \, {\rm cm^{-2}}$).
The total hydrogen column density within a cosmological 
shock is $N_{\rm asy} (1+z)^2 \eta$, where
\begin{equation}
     N_{\rm asy} = (1.6 \times 10^{19} \, {\rm cm^{-2}}) E_{61}^{1/5}
          h_{50}^{6/5} \Omega_{\rm IGM}^{4/5} (1+z_1)^{1/5} \; \; .
\end{equation}
Figure~9 shows how the shocked column density evolves for various 
blastwave parameters and $\Omega_{\rm IGM} = 0.01$.  Behind 
a strong adiabatic cosmological shock, $D_{\rm cl} = 4$ until $t \sim 
\Omega_{\rm IGM}^{-3/2} t_1$ (\S~2.5.3).  Initially, electron collisions 
keep the postshock gas highly ionized.  As the shock slows, photoionization
maintains a relatively high ionization level.  If collisional ionization 
can be neglected, the neutral column behind the shock is
\begin{equation}
     N_{\rm H \, I} \approx (1.7 \times 10^{13} \, {\rm cm^{-2}}) (1+z)^5 
          E_{61}^{1/5} h_{50}^{16/5} \Omega_{\rm IGM}^{9/5} 
          T_4^{-0.8} J_{21}^{-1} (1+z_1)^{1/5} \eta \; \; .
\end{equation}
Figure~10 illustrates $N_{\rm H \, I}(z)$ for a range of blastwave 
parameters and $\Omega_{\rm IGM} = 0.01$.  The lines indicating 
$N_{\rm H \, I}$ are dashed when collisional ionization exceeds
photoionization and solid when photoionization exceeds collisional ionization.
Low-energy blastwaves produce larger neutral columns because they have
smaller postshock temperatures ($T_4 \propto E_{61}^{2/5}$).  Unless
$\Omega_{\rm IGM} \sim 1$, Ly$\alpha$ absorption in adiabatic blastwaves
is much weaker than in Ly$\alpha$ clouds. 

Cooling can potentially magnify the postshock density, boosting the neutral column in the blastwave, but the compression ratios of radiative 
cosmological shocks in the ambient IGM are unlikely to be very large.  
At $z \sim 2$, shocks with $E_0 \lesssim 10^{57} \, {\rm erg}$ can 
cool in a Hubble time in a medium with $\Omega_{\rm IGM} \lesssim 0.03$
(Figure~2).  The requisite shock velocities are $\lesssim 40-50 \, 
{\rm km \, s^{-1}}$, not much larger than the velocity dispersions 
of Ly$\alpha$ clouds.   The widths of intergalactic Ly$\alpha$ lines, 
if thermal, imply temperatures $\sim 3 \times 10^4 \, {\rm K}$, 
consistent with photoelectric heating by quasars.  If the intercloud 
gas is at least as hot, radiative IGM shocks have Mach numbers $\lesssim 3$,
too weak to produce $N_{\rm H I} \sim 10^{14} \, {\rm cm}^{-2}$
when $\Omega_{\rm IGM} \lesssim 0.03$.  Allowing $\Omega_{\rm IGM} \sim 0.1$
requires an IGM temperature $\gtrsim 10^6 \, {\rm K}$ for Ly$\alpha$ 
to remain transparent.  Even in this denser environment, the Mach numbers 
of radiative shocks, with velocities $< 200 \, {\rm km \, s^{-1}}$ 
(\S~2.2), would still be too low to increase $D_{\rm cl}$ sufficiently.

\subsection{Shock-Cloud Interactions}

Since blastwaves in a low-density IGM have insignificant neutral 
columns, Ly$\alpha$ clouds form through some other means.  The porosity 
and metallicity constraints discussed in \S~5 argue against pressure 
confinement of Ly$\alpha$ clouds.  Mechanisms in which interstellar
gas collects in ``minihaloes'' of dark matter (e.g. Rees 1986)
or converges at caustics in the large-scale velocity field 
(e.g. McGill 1990, Cen \etal 1994) appear to be more likely.
Fluctuations in the spatial distribution of collisionless matter 
certainly alter the density field of the IGM, which we have so
far idealized as uniform.  Real intergalactic shocks propagate through 
an inhomogeneous medium; hence, the solutions derived above become accurate 
only after the shock radius exceeds the length scales of non-linear density perturbations.  The remainder of this section briefly examines how 
interactions between cosmological blastwaves and density inhomogeneities 
proceed and suggests that intergalactic blastwaves might amplify the 
neutral column densities of preexisting Ly$\alpha$ clouds.

Shock propagation into an inhomogeneous interstellar medium has been 
analyzed extensively (McKee \& Cowie 1975; Woodward 1976; 
Nittman, Falle, \& Gaskell 1982; Stone \& Norman 1992; 
Klein, McKee, \& Colella 1994).  When a shock of 
velocity $v_s$ encounters a cloud $D_{\rm cl}$ times denser than the 
ambient medium, it drives a shock of velocity $D_{\rm cl}^{-1/2} v_s$ 
into the cloud.  In a cloud-crushing time ($t_{\rm cc} \sim 
D_{\rm cl}^{1/2} r_{\rm cl} v_{\rm cl}$), the internal shock crosses 
the cloud.  Detailed numerical simulations show that the vorticity 
generated in the cloud-shock collision shreds the cloud in a few 
cloud-crushing times (Nittman \etal 1982; Stone \& Norman 1992;
Klein \etal 1994).  
If Ly$\alpha$ clouds are indeed several hundred kpc in size, the cloud 
shredding time approaches a Hubble time for $v_s$ less than a few times
$10^2 \, {\rm km \, s^{-1}}$.  Since the postshock cooling time in the 
cloud is smaller than the ambient cooling time by a factor 
$\sim D_{\rm cl}^{2}$, such long-lived clouds, or fragments thereof, 
might collapse to much higher density before vorticity destroys 
the cloud.  The resulting $N_{\rm H \, I}$ would then increase by 
a factor $\sim v_s^2 / v_{th}^2$, where $v_{th} \sim 30 \, 
{\rm km \, s^{-1}}$ is the thermal velocity dispersion of the 
photoionized cloud after it has cooled to its equilibrium temperature.
Such a shock would impart a velocity $D_{\rm cl}^{-1/2} v_s$ to the
cloud, perhaps carrying it out of its potential well.
Figure 11 illustrates how large $D_{\rm cl}$ must be for shocked clouds 
to cool within a Hubble time.  In the example shown, shocks of $E_0
= 10^{57}$, $10^{59}$, and $10^{61} \, {\rm erg}$ begin propagating
at $z_1 = 4$ in an $\Omega = 1$, $h_{50} = 1$ universe with 
$\Omega_{\rm IGM} = 0.01$.  Clouds with $D_{\rm cl} = 6$ at 
$z \gtrsim 2$, for instance, can cool after encountering shocks 
of $v_s \lesssim 200 \, {\rm km \, s^{-1}}$.  

Computing in detail how cosmological blastwaves interact with developing
gaseous structures in the IGM is beyond the scope of this paper; however, 
the cooling times estimated here indicate that blastwaves could dramatically alter the observable characteristics of preexisting Ly$\alpha$ clouds.
In essence, shocked clouds are pressure-confined, but the pressurizing 
medium does not pervade the universe.  Instead, the pressurized
regions are the interiors of individual blastwaves.  Clouds near blastwave sources would then have systematically higher neutral column densities
than clouds in the ambient IGM.

\section{Summary}

This paper has investigated the behavior of protogalactic and quasar-driven
blastwaves in an expanding universe with $\Omega_{\rm IGM} \ll \Omega_0$.
It presents a new analytical solution for adiabatic blastwaves that applies
until $t \sim t_1 \Omega_{\rm IGM}^{-3/2}$, when self-gravity starts to
compress the postshock gas into a thin shell.  Although two different
timescales ($t$ and $t_1$) enter into the problem, a self-similar solution
still exists.  The similarity of the universe's expansion allows us to define another time coordinate ($\hat{t}$) in which the Sedov-Taylor solution
satisfies an appropriately transformed set of fluid equations if 
$\gamma = 5/3$.  Cosmological blastwaves thus begin as Sedov-Taylor
blastwaves and approach a constant comoving radius as $t$ grows large.
While $t \ll t_1 \Omega_{\rm IGM}^{-3/2}$, the gas density immediately
behind the shock remains equal to four times the ambient density, so 
cosmological shocks do not cool until they are moving quite slowly.

Applying this solution to protogalactic starbursts and quasars shows that
small-scale explosions ($\lesssim 10^{57} \rm {\rm erg}$) are probably
more important than large-scale explosions ($\gtrsim 10^{58} \, {\rm erg}$)
in determining the state of the IGM and its Ly$\alpha$ clouds.  
Large-scale explosions are quite inefficient at ionizing the IGM.
Cosmological effects rob most of the energy from large-scale 
blastwaves before these shocks slow to speeds that ionize efficiently
($\lesssim 100 \, {\rm km \, s^{-1}}$).  In addition, since the comoving
volume per unit energy of a blastwave is $\propto E_0^{-2/5}$,
small-scale explosions can fill the IGM much more easily.  Quasar-driven
blastwaves are too rare to fill the IGM by $z \sim 4$ without violating
the $y$-parameter limit from COBE.  Large-scale protogalactic
blastwaves that filled the universe by such an early time would raise 
the mean metallicity of the IGM well above the values observed in 
Ly$\alpha$ clouds at $z \sim 2-3$.

Contrary to some suggestions that Ly$\alpha$ clouds might be shells
of gas shocked by cosmological blastwaves, the solution derived here
shows that the postshock gas never becomes dense enough to produce 
$N_{\rm H \, I} \gtrsim 10^{13} \, {\rm cm^{-3}}$ when $\Omega_{\rm IGM}
\lesssim 0.03$.  Even if the blastwave has slowed enough to cool, the 
compression of the ambient IGM is still too small to raise $N_{\rm H \, I}$
to the levels observed in Ly$\alpha$ clouds.  Blasts that pass through 
overdense regions of the IGM, such as preexisting Ly$\alpha$ clouds,
might still produce observable signatures.  Compression and cooling can
dramatically amplify the neutral fraction in a shocked cloud, so
clouds engulfed by blastwaves could have systematically higher values
of $N_{\rm H \, I}$.  This effect would increase the probability of finding
clouds near blastwave sources.

\vspace*{2.0em}

During the development of this paper, conversations with Steve Balbus, 
Megan Donahue, Mike Fall, Piero Madau, and Sterl Phinney were particularly
helpful.  The author gratefully acknowledges the hospitality of the Virginia 
Institute for Theoretical Astrophysics during the early stages of 
this project.  The work presented here was supported by NASA 
through grant number HF-1054.01-93 from the Space Telescope 
Science Institute, which is operated by the Association of 
Universities for Research in Astronomy, Inc., under
NASA contract NAS5-26555. \\

\pagebreak

\pagebreak

\begin{figure}
\plotone{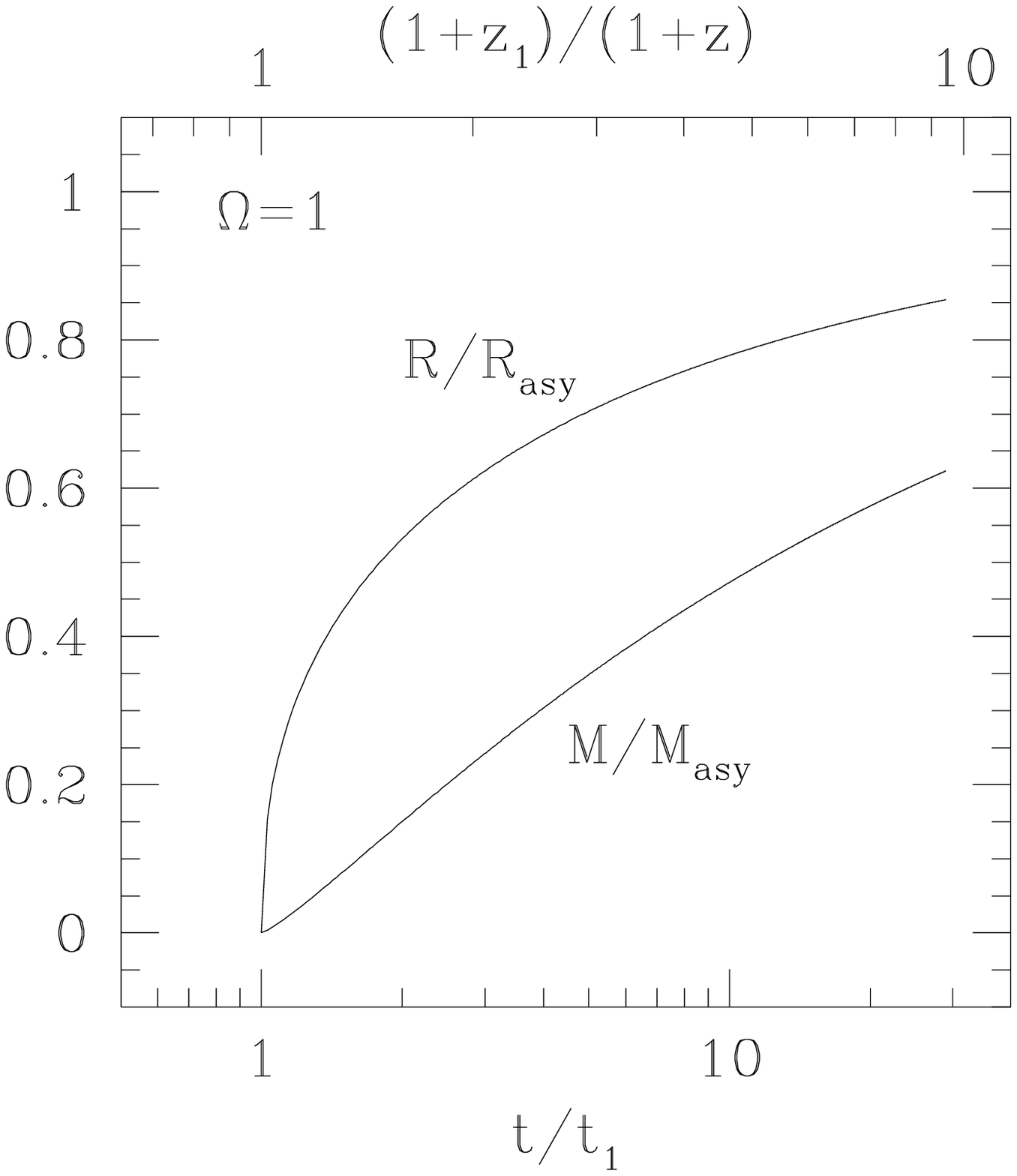}
\caption[]{Growth of an adiabatic cosmological blastwave when
$\Omega _{\rm IGM} \ll \Omega_0 = 1$.  This figure illustrates how quickly
the radius and mass of an adiabatic cosmological blastwave arising from
an intergalactic explosion at cosmic time $t_1$ approach their asymptotic 
values.  At cosmic time $t$, the shock radius has reached a fraction 
$R/R_{\rm asy}$ of its asymptotic comoving value and contains a fraction
$M/M_{\rm asy}$ of the mass it will ultimately encompass.}
\end{figure}

\begin{figure}
\plotone{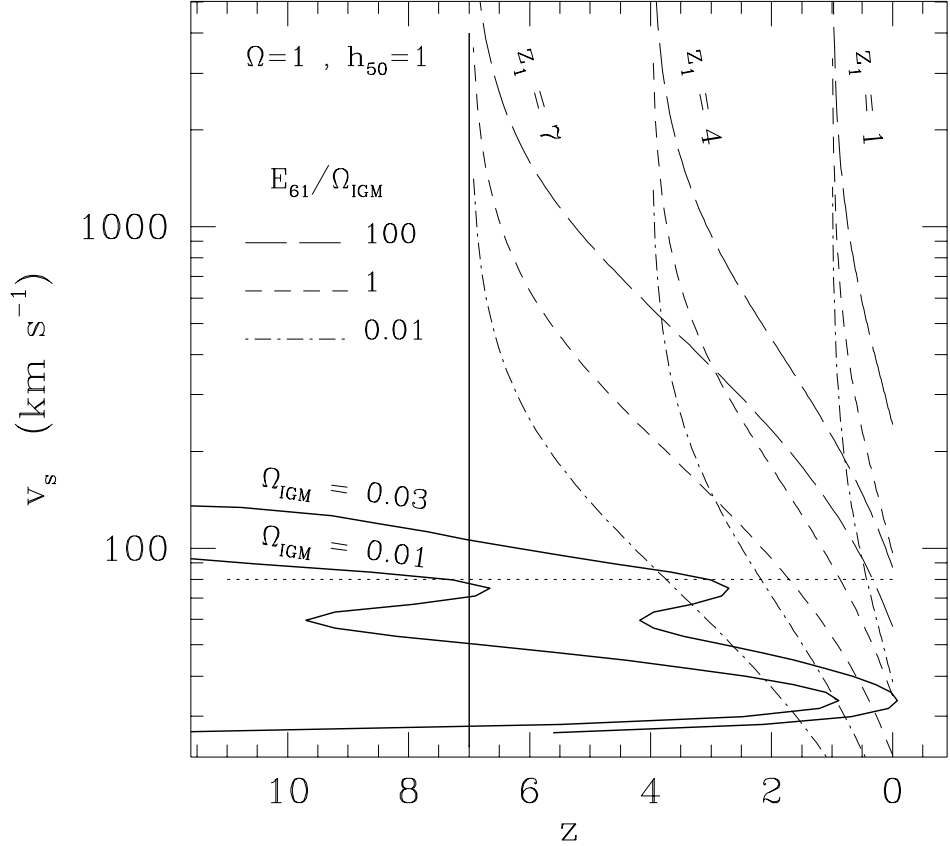}
\caption[]{Shock velocities of adiabatic cosmological blastwaves.
The velocity $v_s(z)$ of the adiabatic shock driven into the IGM when
an energy $(10^{61} \, {\rm erg}) E_{61}$ is introduced at redshift
$z_1$ depends on $z_1$, $\Omega$, and the value of $E_{61} h_{50} 
\Omega_{\rm IGM}^{-1}$.  Here we show how $v_s$ varies with $z$ for
several values of $E_{61} \Omega_{\rm IGM}^{-1}$ when $h_{50} = 1$,
$\Omega = 1$, and $\Omega_{\rm IGM} \ll 1$.  Curves corresponding to
shocks beginning at three different initial redshifts ($z_1 = 1, 4, 
7$) are shown.  The solid lines indicate where cooling begins to violate 
the assumption that the shocks are adiabatic.  The vertical line at 
$z = 7$ shows where Compton cooling against the microwave background 
becomes important, and the two lines across the bottom show where 
radiative cooling becomes important for $\Omega_{\rm IGM} = 0.01$ and 
0.03, assuming an IGM metallicity of 0.01 times solar (radiative 
cooling functions from Sutherland \& Dopita 1993).  The horizontal 
dashed line shows where ionization losses would become important
if the ambient medium were neutral.  Since the $v_s(z)$ curves lie
primarily above and to the right of the solid lines, cosmological
shocks at $z<7$ are adiabatic for a wide range of parameters.}
\end{figure}

\begin{figure}
\plotone{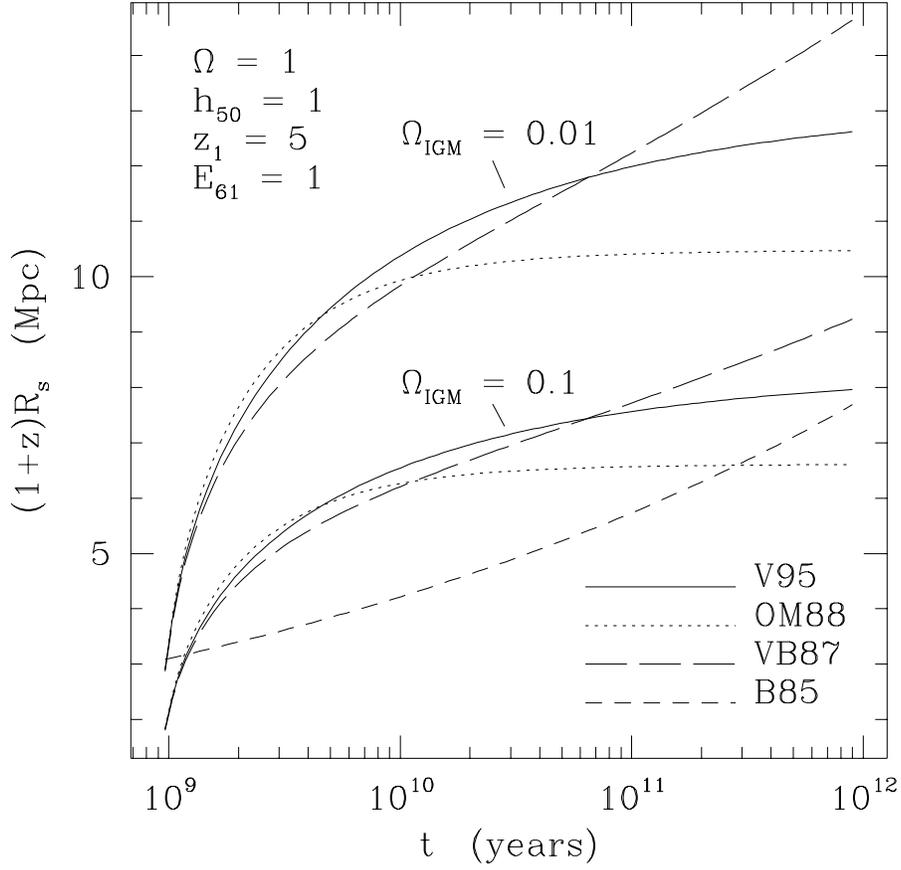}
\caption[]{Approximate solutions for adiabatic cosmological blastwaves 
when $\Omega_{\rm IGM} \ll \Omega_0 = 1$.  Solid lines (V95) show the exact 
solutions from \S~2.1 for $R_s(t)$ in comoving coordinates, given a
$10^{61} \, {\rm erg}$ blastwave starting at $z_1 = 5$ in intergalactic
media with $\Omega_{\rm IGM} = 0.1$ and $\Omega_{\rm IGM} = 0.01$.
These solutions become invalid when $t \sim t_1 \Omega_{\rm IGM}^{-3/2}$.
Dotted lines (OM88) show the approximate solutions derived by
Ostriker \& McKee (1988) for these same parameters.  The long-dashed 
line (VB87) gives the approximate fit of Vishniac \& Bust (1987)
to numerical models.  The short-dashed line (B85) gives the 
self-similar solution from Bertschinger (1985), valid when 
$t \gg t_1 \Omega_{\rm IGM}^{-3/2}$ and the shocked IGM comoves 
with a collisionless shell that has responded to the blastwave.}
\end{figure}

\begin{figure}
\plotone{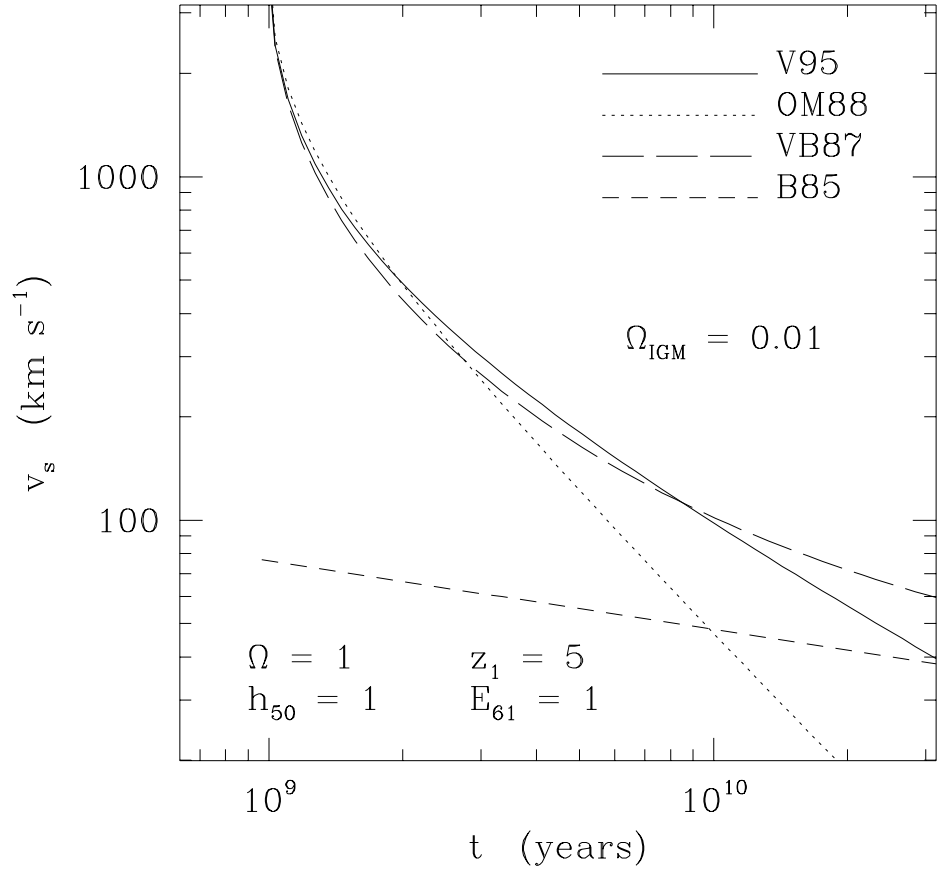}
\caption[]{Approximate shock velocities.  The lines in this figure
trace the shock velocities given by the approximate blastwave solutions
in Figure~3.}
\end{figure}

\begin{figure}
\plotone{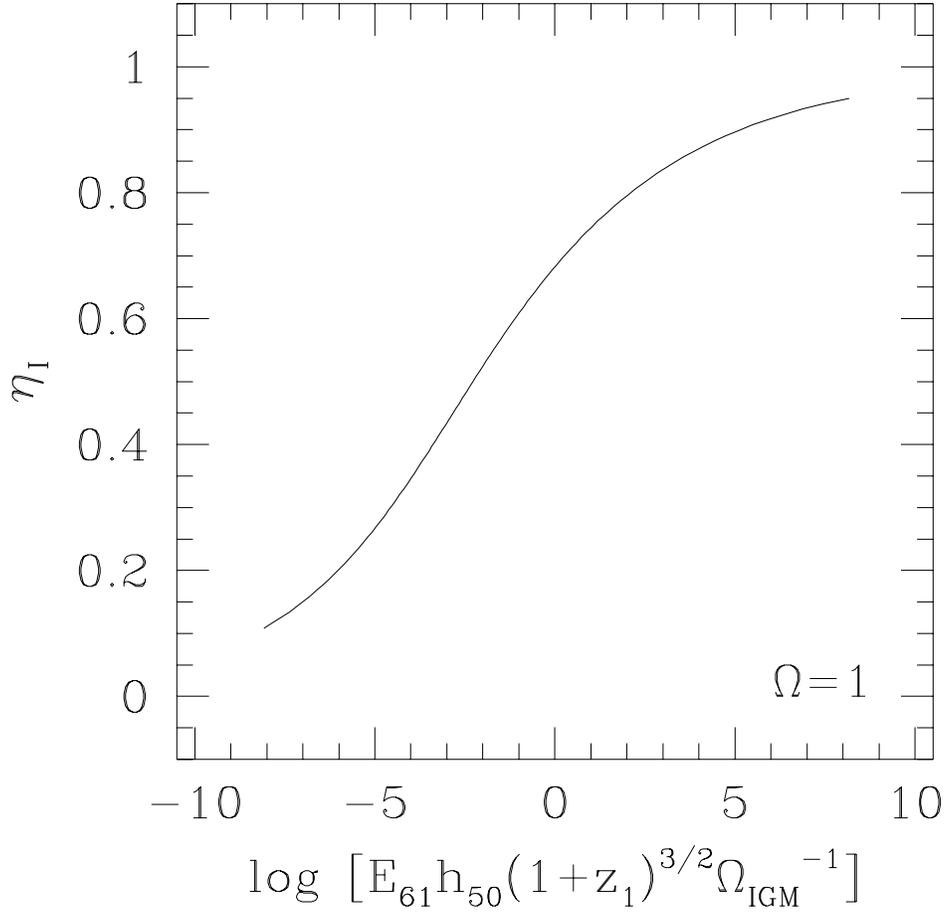}
\caption[]{Dimensionless comoving radius at which shock ionization
becomes ineffective ($\Omega = 1$).  The quantity $\eta_I$ is the value 
of $\eta \equiv \hat{R}_s / \hat{R}_{\rm asy}$ at which the shock 
velocity drops below $50 \, {\rm km \, s^{-1}}$.  Beyond this point, 
the shock no longer provides enough energy to ionize the incoming hydrogen.  When $E_{61}$ is large, $\eta_I$ is of order unity, and cosmological 
effects have begun to govern the growth of the blastwave.}
\end{figure}

\begin{figure}
\plotone{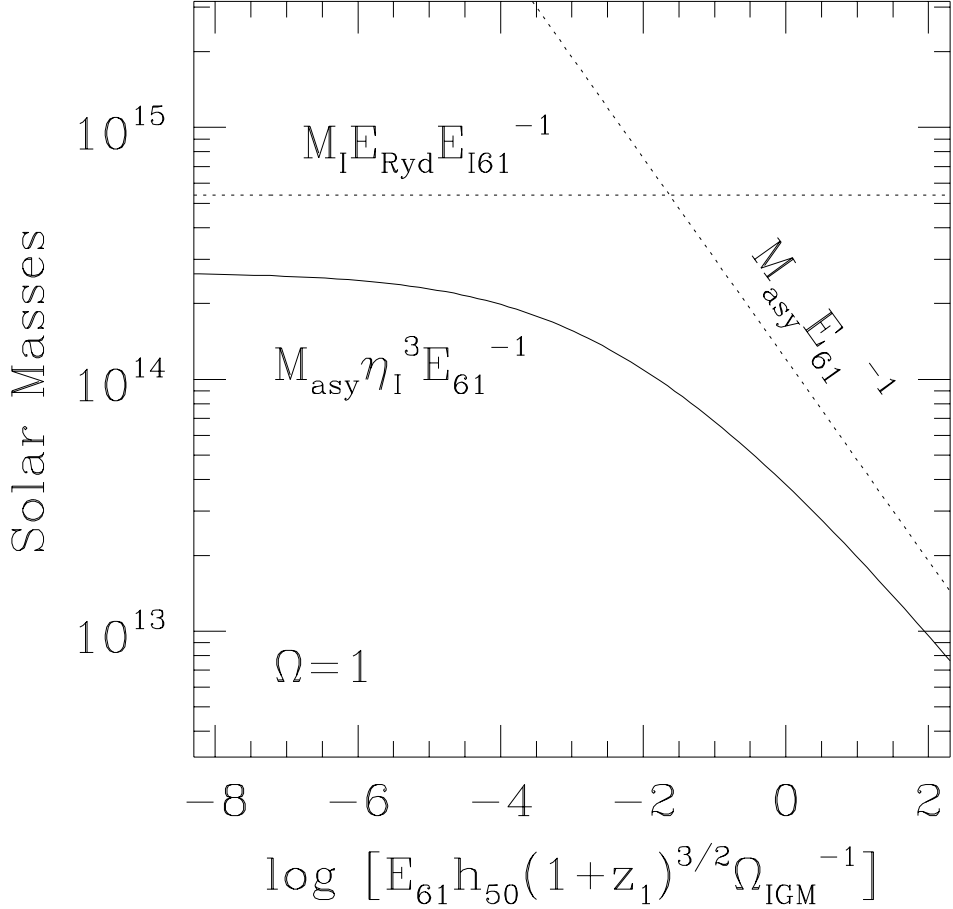}
\caption[]{Total ionized mass when $\Omega = 1$.  The horizontal 
dotted line gives the total mass ($M_I$) ionized by $(10^{61} \, 
{\rm erg}) E_{I61}$ in ionizing photons with mean energy 
$(13.6 \, {\rm eV}) E_{\rm Ryd}$.  An adiabatic blastwave initially 
containing $(10^{61} \, {\rm erg}) E_{61}$ ultimately encompasses a 
mass $M_{\rm asy}$.  The diagonal dotted line shows how this
asymptotic mass varies with the parameter combination $E_{61} h_{50} 
(1+z_1)^{3/2} \Omega_{\rm IGM}^{-1}$ in a flat universe.  
The solid line gives the amount of mass ($M_{\rm asy} \eta_I^3$) 
within an adiabatic blastwave
when the shock ceases to ionize the incoming gas efficiently.
The maximum amount of collisionally ionized mass per unit input energy 
drops with increasing energy because high-energy blastwaves
transfer most of their energy to the collisonless component before
ionization losses become important.  Cosmological blastwaves do not 
ionize the IGM unless their sources emit relatively few ionizing 
photons into intergalactic space.}
\end{figure}

\begin{figure}
\plotone{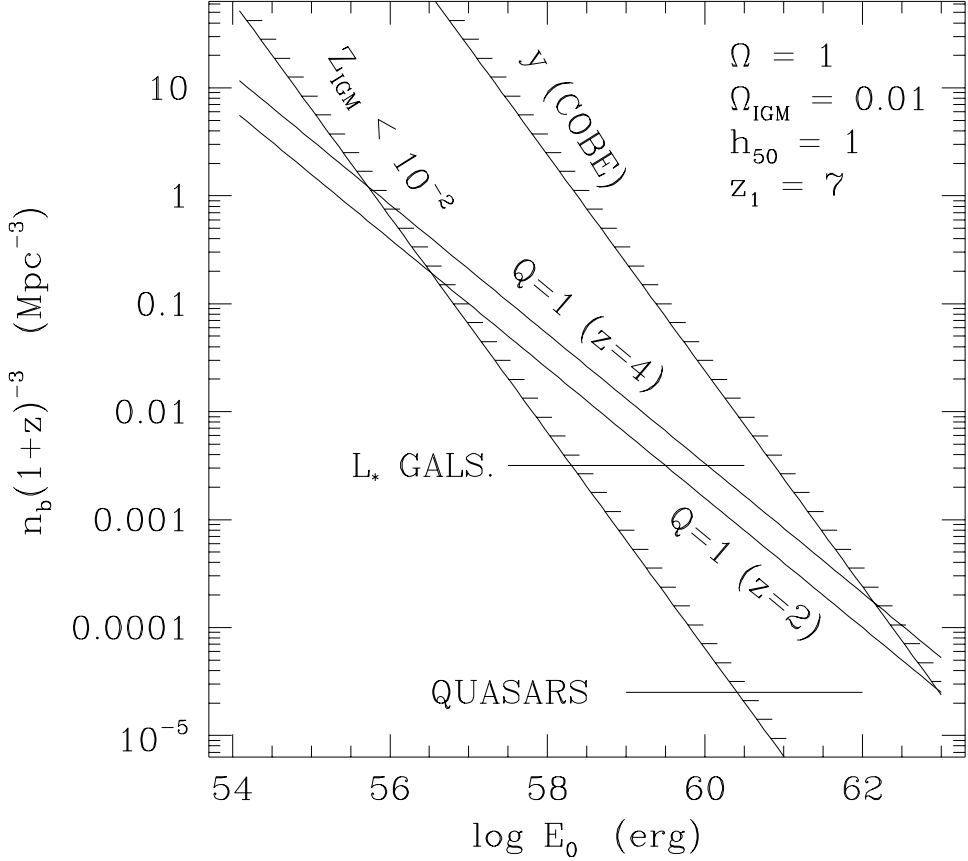}
\caption[]{Constraints on porosity.  COBE limits on the $y$-distortion
of the microwave background constrain the thermal energy density of the 
IGM at high redshift.  The product of the energy per blastwave, $E_0$, and
the comoving number density of blastwave sources, $n_b (1+z)^{-3}$, must 
therefore lie below the line marked $y$~(COBE), if $\Omega = 1$, 
$h_{50} = 1$, and $z_1 = 7$.  Blastwaves driven into the IGM by multiple 
supernovae introduce metals along with thermal energy.  If the metallicities 
of Ly$\alpha$ clouds ($\lesssim 10^{-2}$ times solar) are similar to the 
mean metallicity of the IGM, the energy density injected by supernovae 
must be significantly lower than the COBE limit.  The line marked 
$Z_{\rm IGM} < 10^{-2}$ indicates the value of $E_0 n_b (1+z)^{-3}$ that 
would yield an IGM metallicity of $10^{-2}$ times solar, assuming each 
supernova produces $1 \, M_\odot$ of metals and $\Omega_{\rm IGM} = 0.01$.  
The lines labeled $Q=1$ trace the locus of blastwave parameters that give 
$Q=1$ at $z=4$ and $z=2$ for blastwaves that begin propagating at $z_1 = 7$.
(See Fig. 8 caption for additional information.)}
\end{figure}

\begin{figure}
\plotone{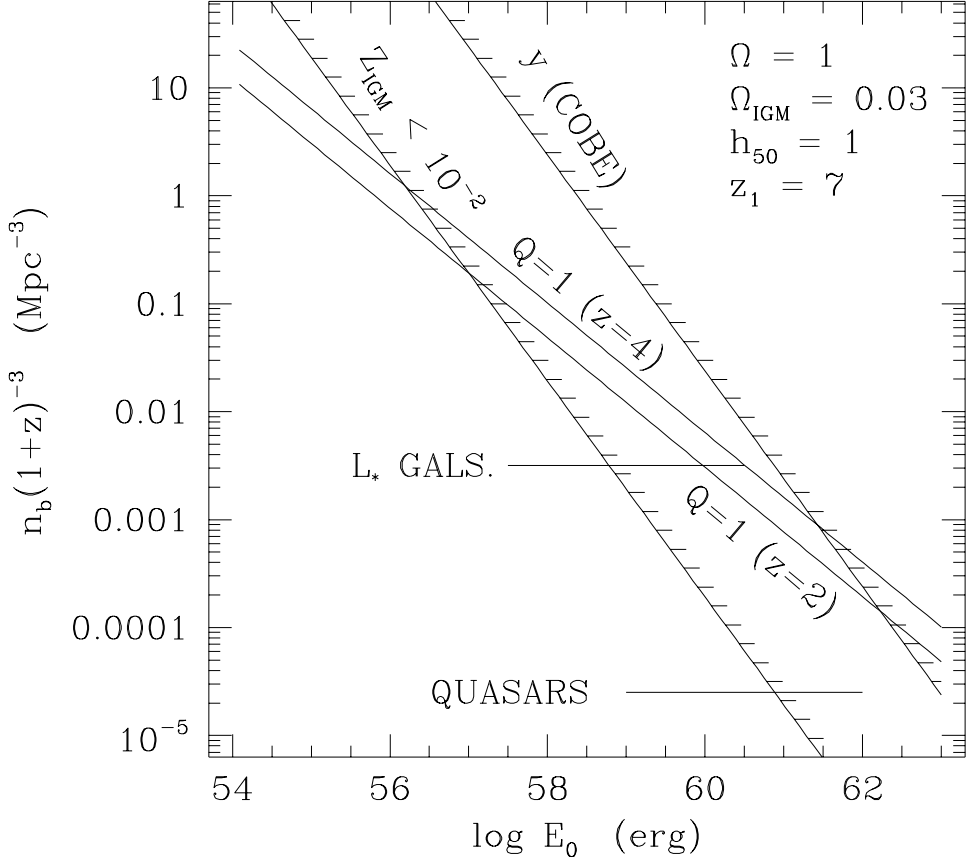}
\caption[]{Constraints on porosity.  Same as Figure~7, except 
$\Omega_{\rm IGM} = 0.03$.
The metallicity constraint allows $Q \gg 1$ at $z > 2$ only if blastwave
sources are much more numerous than $L_*$ galaxies and the energy per
blastwave is $\ll 10^{57} \, {\rm erg}$.  Quasar blastwaves could be
metal-free and need not satisfy the metallicity constraint;  however,
the COBE constraint rules out $Q(z > 2) \gg 1$ in quasar blastwaves
if $\Omega_{\rm IGM} \geq 10^{-2}$.  The line marked QUASARS
indicates the comoving number density of quasars measured at $z = 2$ 
by Boyle (1993), multiplied by a factor of 10 to account for the quasar
duty cycle (quasar lifetime $\sim 10^8 \, {\rm yr}$ divided by $t
\sim 10^9 \, {\rm yr}$ at $z \sim 5$).  Blastwaves driven by objects 
this uncommon cannot fill the universe by $z > 2$, while still satisfying 
the COBE constraint, regardless of their energy output.}
\end{figure}

\begin{figure}
\plotone{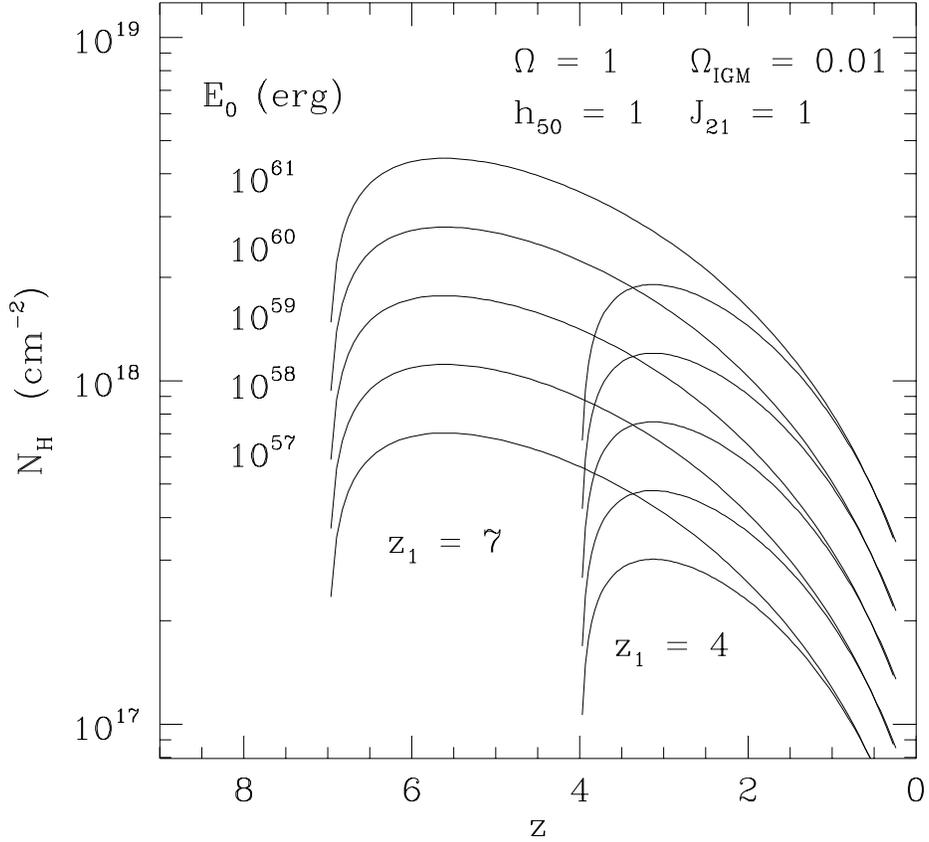}
\caption[]{Postshock H column density.  This figure shows how the 
column density of hydrogen behind a spherical cosmological shock 
changes with time in a uniform IGM.  The curves indicate $N_{\rm H}$
in blasts beginning at redshifts $z_1 = 4$ and 7 with initial energies
in the range $10^{57} - 10^{61} \, {\rm erg}$.  We assume $\Omega_{\rm
IGM} = 0.01$, $h_{50} = 1$, and $\Omega = 1$.  Note $N_{\rm H} \propto
\Omega_{\rm IGM}^{4/5} h_{50}^{6/5}$.  At first, $N_{\rm H}$ grows.
Later, as the shock begins to merge with the Hubble flow, $N_{\rm H}$
drops.}
\end{figure}

\begin{figure}
\plotone{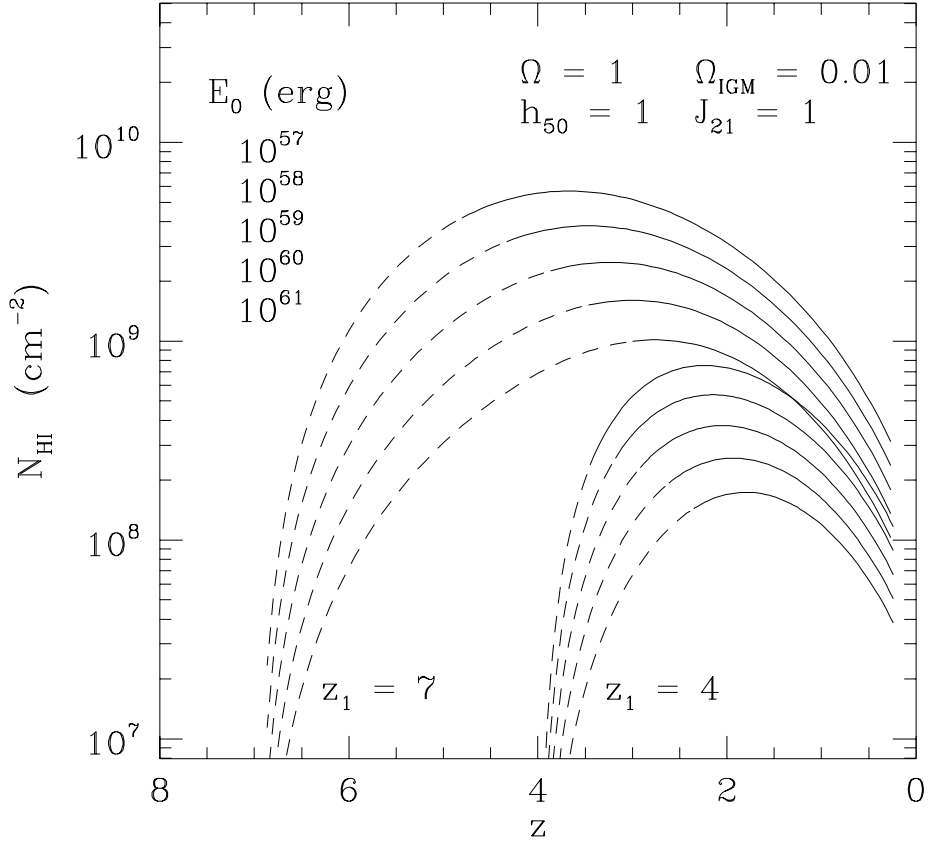}
\caption[]{Postshock H~I column density.  This figure gives the
neutral postshock column densities for the same blastwaves described
in Figure~9.   The shocks are presumed to be adiabatic (not true for 
$E_0 \sim 10^{57} \, {\rm erg}$ at $z < 1.5$) and strong enough for 
the postshock density to be four times the IGM density.
Collisional ionization governs postshock ionization
balance at early times, indicated by the dashed lines.  Later, 
photoionization ($J_{21} = 1$) overtakes collisional ionization.  
The transition from dashed to solid lines occurs where photoionization 
and collisional ionization are equally important.  
Low energy blastwaves have higher $N_{\rm H \, I}$
because their postshock temperatures are lower, allowing recombination
to act more rapidly.  Note $N_{\rm H \, I} \propto \Omega_{\rm IGM}^{9/5}
h_{50}^{16/5}$. }
\end{figure}

\begin{figure}
\plotone{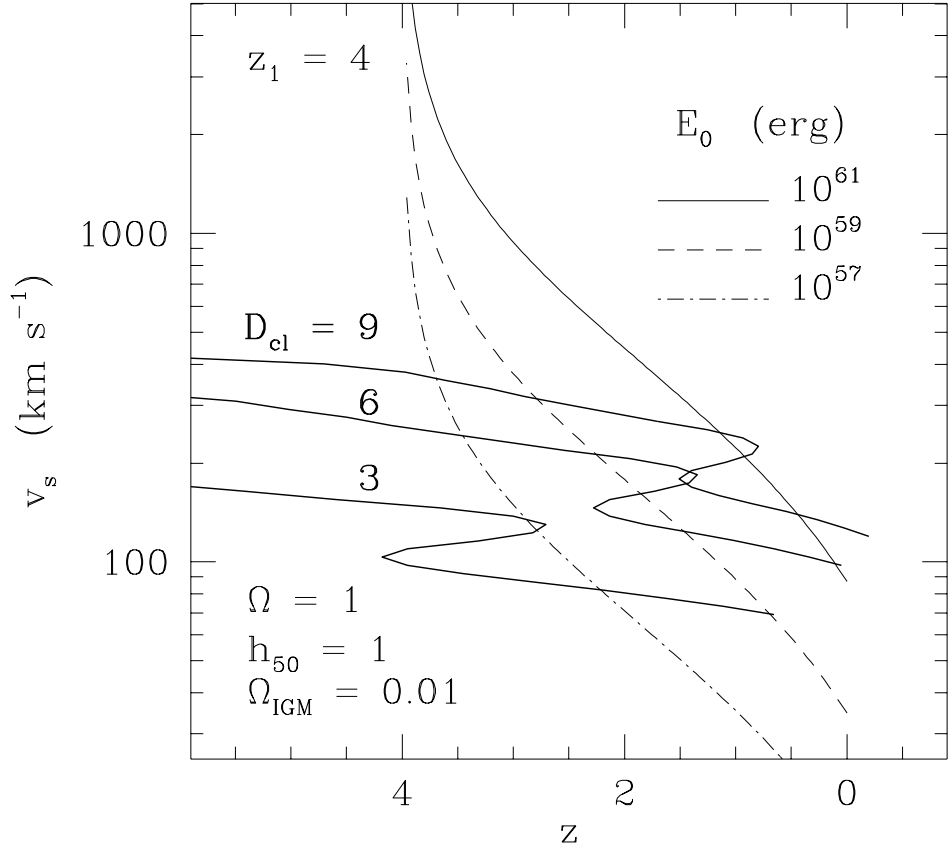}
\caption[]{Conditions under which shocked clouds cool.  Adiabatic
shocks that encounter clouds $D_{\rm cl}$ times denser than the ambient 
IGM can drive radiative shocks into these clouds if $D_{\rm cl}$ is 
sufficiently large.  The heavy solid lines indicate the shock speeds
$v_s$ through the ambient medium below which shocked clouds cool in a 
Hubble time at various redshifts $z$.  Here we show cooling loci for
$\Omega = 1$, $h_{50} = 1$, and $\Omega_{\rm IGM} = 0.01$.  For reference,
we also show $v_s (z)$ trajectories for blastwaves beginning at $z_1 = 4$
with $E_0 = 10^{57}$, $10^{59}$, and $10^{61} \, {\rm erg}$.  
Because the postshock cooling time is roughly proportional to 
$D_{\rm cl}^{-2}$, clouds of modest overdensities can cool much more 
easily than the ambient IGM.  Thus, cosmological blastwaves can 
potentially amplify the neutral column densities
of preexisting Ly$\alpha$ clouds by a large factor. }
\end{figure}

\end{document}